\documentclass[aps,floats,superscriptaddress,twocolumn,pre,showpacs]{revtex4-1}

\usepackage{graphicx,epsfig}% Include figure files
\usepackage{graphics,dcolumn,bm,epic, eepic,fleqn,float}
\usepackage{amssymb,amsmath,amsfonts,multirow,rotate,color}
\usepackage{soul}
\usepackage{wasysym}
\usepackage{dcolumn}
\usepackage{bm}

\definecolor{Myorange}{cmyk}{0,0.42,1,0}

\newcommand{\lay}[1]{^{[#1]}}

\def\Erdos{Erd\"os}

\begin{document}

\title{Structural measures for multiplex networks}

\author {Federico Battiston}
\affiliation{School of Mathematical Sciences, Queen Mary University of London, 
London E1 4NS, United Kingdom}   

\author {Vincenzo Nicosia}
\affiliation{School of Mathematical Sciences, Queen Mary University of London, 
London E1 4NS, United Kingdom}  
\affiliation{Laboratorio sui Sistemi Complessi, Scuola Superiore di Catania, 
I-95123 Catania, Italy}

\author{Vito Latora}
\affiliation{School of Mathematical Sciences, Queen Mary University of London, 
London E1 4NS, United Kingdom}  
\affiliation{Laboratorio sui Sistemi Complessi, Scuola Superiore di Catania, 
I-95123 Catania, Italy}
\affiliation{Dipartimento di Fisica ed Astronomia, Universit\`a di Catania and INFN, I-95123 Catania, Italy}

\date{\today}
%ABSTRACT
\begin{center}
\begin{abstract}
Many real-world complex systems consist of a set of elementary units
connected by relationships of different kinds. All such systems are
better described in terms of multiplex networks, where the links at
each layer represent a different type of interaction between the same
set of nodes, rather than in terms of (single-layer) networks.  In
this paper we present a general framework to describe and study
multiplex networks, whose links are either unweighted or weighted.  In
particular we propose a series of measures to characterize the
multiplexicity of the systems in terms of: {\em i)} basic node and
link properties such as the node degree, and the edge overlap and
reinforcement, {\em ii)} local properties such as the clustering
coefficient and the transitivity, {\em iii)} global properties related
to the navigability of the multiplex across the different layers.  The
measures we introduce are validated on a genuine multiplex data set of
Indonesian terrorists, where information among 78 individuals are
recorded with respect to mutual trust, common operations, exchanged
communications and business relationships.
\end{abstract}
\end{center}

\maketitle

%INTRODUCTION
\section{\textbf{INTRODUCTION}}

Much work has been done in the last decades in investigating and
characterizing the structure and dynamics of complex systems.
Many of these systems have been proven to be
  successfully described as a network whose nodes represent the
  different basic units of the system, and whose links represent the
  interactions/relationships among the units
  \cite{Strogatz2001,Barabasi2002rev,Newman2003rev,Boccaletti06}. The
  standard approach to network description of complex systems consists
  of studying the graph resulting from the aggregation of all the
  links observed between a certain set of elementary units. However,
  such aggregation procedure might in general discard important
  information about the structure and function of the original system,
  since in many cases the basic constituents of a system might be
  connected through a variety of relationships which differ for
  relevance and meaning~\cite{Kurant, Buldyrev}.  For instance, the
same set of individuals in a social system can be connected through
friendship, collaboration, kinship, communication, commercial and
co-location relationships, just to name some of them, while in complex
multi-modal transportation systems, which are typical of large
metropolitan areas, a set of locations might be reached in several
different ways, e.g. using bus, underground, suburban rail, riverboat
networks and the like.  In these systems, each type of interaction has
associated a given relevance, importance, cost, distance or meaning,
so that treating all the links as being equivalent results into losing
a lot of important information. A better description of such systems
is in terms of \textit{multiplex networks}, i.e. networks where each
node appears in a set of different layers, and each layer describes
all the edges of a given type.

Recently, a considerable amount of effort has been devoted to the
characterization and modeling of multiplex networks, with the aim of
creating a consistent mathematical framework to study, understand and
reproduce the structure of these systems. A number of measures have
been proposed in the context of real-world multiplex networks such as
air transportation systems~\cite{Cardillo} and massive multiplayer
online games~\cite{Szell}. Some other works are pointing towards a
statistical mechanics formulation of multiplex
networks~\cite{bianconi}, to the extension of classical network
metrics to the case of multiplexes~\cite{sole,DeDomenico2013} and to
model the growth of systems of this kind~\cite{growing}. Finally,
another active research direction is that of characterizing the
dynamics and the emergent properties of multi-layer systems,
especially with respect to epidemic~\cite{Saumell} and information
spreading~\cite{Cozzo, Min}, cooperation~\cite{Gomez-Gardenes},
synchronization~\cite{Nicosia}, diffusion processes~\cite{Gomez} and
random walks on multiplex networks~\cite{dedomenico}. A review of
recent papers in this field can be found in Ref.~\cite{Kivela2013}.

In this article we introduce a set of basic metrics to characterize
the structural properties of multiplex networks, including their
degree distributions, edge overlap, node clustering, spectral
centrality, configuration of shortest paths, betweenness and closeness
centrality. In particular, we focus on the quantification of the
participation of single nodes to the structure of each layer, and of
the importance of each node for the overall efficiency of the
multiplex network, in terms of node reachability and triadic
closure. All the proposed measures are tested and validated on a
genuinely multiplex real-world data set, the Top Noordin Terrorist
Network, which includes detailed information about four different
features, namely mutual trust, common operations, exchanged
communications and business involvement of 78 Indonesian
terrorists. Thanks to its peculiar structure, this system can be
naturally modeled as a four-layer multiplex.  We show that, in this
particular data set, one of the four layers, namely the trust layer,
acts as a driver for the others, since the conditional probability for
two terrorists to communicate or to participate to the same operation
clearly depends on the strength of their mutual trust
relationship. This result can be explained in terms of social
reinforcement, and reveals important details about the overall
dynamics of edge formation and strengthening in the multiplex.  We
believe that the measures proposed hereby will have wide applicability
to larger multiplexes in several different domains.

The article is organized in the following way. In
Section~\ref{section:formalism} we discuss the various levels at which
we can describe a multiplex network. We introduce the aggregated
topological matrix, the overlapping and the weighted overlapping
matrix, and the vector of adjacency matrices $\bm{A}$, which provides
a complete description of the multiplex network. We also discuss basic
metrics, including node degree and edge overlap.  In
Section~\ref{section:indonesian} we introduce the multi-layer system
under study, a multiplex network with four layers describing the
interactions among $78$ Indonesian terrorists.  In
Section~\ref{section:node} we compare the different measures of node
degree on the network under study and we introduce metrics to describe
how the links of a node are distributed over the various layers.  In
Section~\ref{section:overlap} we quantify the edge overlap and we
discuss a mechanism of social reinforcement present in the network of
terrorists.  In Section~\ref{section:clustering} we generalize the
concepts of clustering and transitivity to the case of multiplex
networks, considering the possibility of triangles with links in
different layers.  In Section~\ref{section:paths} we investigate the
number of shortest paths which make use of links in different layers.
In Section~\ref{section:centrality} we propose a simple extension of
spectral centrality to networks with multiple layers. Finally, in
Section~\ref{section:conclusions} we present our conclusions.

%AGGREGATED AND OVERLAPPING
\section{\textbf{General formalism}}
\label{section:formalism}
Consider a complex system involving multiple kinds of relations among
its basic units. When it is possible to distinguish the nature of the
ties, an effective approach to describe the system consists in
embedding the edges in different layers according to their type. This
is the starting point of multiplex networks analysis.

In this section we propose a comprehensive approach and a coherent
notation for the study of systems composed of $N$ nodes and $M$
layers, with ties in each layer being undirected and either
unweighted or weighted~\cite{note_undirected}. Our framework does not
fit, instead, the case of a multiplex of multiplexes, i.e. a system in
which each layer is composed by a number of sub-layers (which may in
turn be composed by several sub-sub-layers, and so on.), but might be
easily extended to encompass this case. 

We consider first a system composed of $N$ nodes and $M$ unweighted
layers, and we extend the notation to the case of weighted layers
afterwards. We can associate to each layer $\alpha,
\alpha=1,\ldots,M,$ an adjacency matrix
$A^{[\alpha]}=\{a_{ij}^{[\alpha]}\}$, where $a_{ij}^{[\alpha]}=1$ if
node $i$ and node $j$ are connected through a link on layer $\alpha$,
so that each of the $M$ layers is an unweighted network. Such a
multiplex system is completely specified by the vector of the
adjacency matrices of the $M$ layers
\begin{equation}
{\bm A}=\{A^{[1]}, ... , A^{[M]}\}.
\end{equation}
We define the degree of a node $i$ on a given layer as
$k_i^{[\alpha]}=\sum_j a_{ij}^{[\alpha]}$, from which follows that $0
\le k_i^{[\alpha]} \le N-1$ $\forall i, \forall \alpha$. Consequently,
the degree of node $i$ in a multiplex network is the vector
\begin{equation}
 {\bm k_i} = (k_i^{[1]}, ... ,k_i^{[M]}),\quad i=1,\ldots,N
\end{equation}
We have $\sum_i k_i^{[\alpha]} = 2K^{[\alpha]}$, where $K^{[\alpha]}$
is the total number of links on layer $\alpha$.  As for single-layer
networks, we use lowercase letters to denote node properties and
capital letters for properties obtained by summing over the nodes or
the edges, either at the level of single layer or at the level of the
whole system.

Vectorial variables, such as ${\bm A}$ and ${ \bm k_i}$, are necessary
to properly store all the richness of multiplex networks. However, it
is also useful to define aggregated adjacency matrices (in which we
disregard the fact that the links belongs to different layers) to be
used as a term of comparison. As part of the goal of this paper, we
will show that aggregated matrices and the corresponding aggregated
measures with which one may be tempted to analyze the multi-layer
structure have limited potential and often fail in detecting the key
structural features of a multiplex network. We define the aggregated
topological adjacency matrix $\mathcal A=\{a_{ij}\}$ of a multiplex
network, where
\begin{equation}
a_{ij} = \begin{cases}1 & \text{if} \>\>\>\exists\> \alpha:
  a\lay{\alpha}_{ij}=1 \\ 0 & \text{otherwise} \end{cases}
\end{equation}
This is the adjacency matrix of the unweighted network obtained from
the multi-layer structure joining all pairs of nodes $i$ and $j$ which
are connected by an edge in at least one layer of the multiplex
network, and neglecting the possible existence of multi-ties between a
pair of nodes and the nature of each tie as well.  For the degree of
node $i$ on the aggregated topological network, we have
\begin{equation} 
k_i=\sum_j a_{ij}.
\end{equation}
Summing $k_i$ over all elements of the system, we obtain 
\begin{equation}
\sum_i k_i = 2 K,
\end{equation}
where $K$ is the total number of links (also called the size) of the
aggregated topological network. Matrix $\mathcal A$ describes a
single-layer binary network which can be studied using the
well-established set of measures defined for single-layer networks. As
we will show in the following sections, this representation turns out
to be very simplistic and often insufficient to unveil the key
features of multi-layer systems.
A basic feature which is lost in the topological aggregated matrix is
that in multiplex systems the same pair of nodes can be connected by
ties of different kinds. 

We introduce the edge overlap of edge $i-j$ between two layers
$\alpha$ and $\alpha'$ as:
\begin{equation}
o_{ij}^{[\alpha,\alpha']} = a_{ij}^{[\alpha]} + a_{ij}^{[\alpha']},
\end{equation}
and the edge overlap of edge $i-j$ as: 
\begin{equation}
o_{ij} = \sum_{\alpha}a_{ij}^{[\alpha]}.
\label{edge_overlap}
\end{equation}
By definition, we have $0 \le o_{ij} \le M$ $\forall i,j$. We can now
define the aggregated overlapping adjacency matrix $\mathcal
O=\{o_{ij}\}$. Matrix $\mathcal O$ is not different from a standard
weighted adjacency matrix of a single-layer network. Even though the
overlapping matrix $\mathcal O$ has a richer structure compared to the
purely topological matrix $\mathcal A$, in this paper we show that
also this matrix eventually fails in featuring a number of basic
structural properties of multiplex networks. In fact, although the
information about the total number of connections (at different
layers) between each pair of nodes is preserved, the loss of knowledge
in identifying the nature of each tie (which is instead conserved in a
vectorial variable such as ${\bm A}$) will often make $\mathcal O$
insufficient to catch important characteristics of multi-structured
systems.

Based on the edge overlap $o_{ij}$, we can also define the overlapping
degree of node $i$ as:
\begin{equation}
o_i= \sum_{j}o_{ij}=\sum_{\alpha}k_i^{[\alpha]}.
\end{equation}
with $o_i \ge k_i$. Slightly different measures of node overlapping
where defined in~\cite{bianconi}. Notice that the overlapping degree
$o_i$ represents the correct factor to normalize the components of the
degree vector ${\bm k_i}$. In fact, we have $(1/o_i)\sum_{\alpha}
k_i^{[\alpha]}=1$.  Summing $o_i$ over all elements of the system, we
obtain:
\begin{equation}
\sum_i o_i = \sum_{\alpha} \sum_i k_i^{[\alpha]}= 2 \sum_{\alpha}K^{[\alpha]}=2O,
\end{equation}
where $O$ is the size of the overlapping network.

We now consider the case of a multiplex network composed of weighted
layers. In such a case, for all the connected pairs of nodes $i$ and
$j$ on each layer $\alpha$ of the multiplex, we have a positive real
number $w_{ij}^{[\alpha]}$, namely the weight of the link $i-j$ at
layer $\alpha$.  A weighted multi-layer network is completely
specified by the vector of its weighted adjacency matrices
$\bm{W}=\{W^{[1]}, ... , W^{[M]}$\}, with
$W^{[\alpha]}=\{w_{ij}^{[\alpha]}\}$.  In analogy with the case of
unweighted layers, also for weighted layers we can define the
aggregated topological adjacency matrix $\mathcal A = \{a_{ij}\}$,
where
\begin{equation}
 a_{ij} = \begin{cases}1 & \text{if} \>\>\>\exists\> \alpha:
   w\lay{\alpha}_{ij}>0 \\ 0 & \text{otherwise.} \end{cases}
\end{equation} 
We can now extend all the previously introduced measures to the case
of weighted multiplexes. 
We define the strength of node $i$ on layer $\alpha$ as $s_i^{[\alpha]}=\sum_j w_{ij}^{[\alpha]}$.
Similarly to the unweighted case, the strength of node $i$ can be represented as a vector
\begin{equation}
\bm{s}_i = (s_i^{[1]}, ... ,s_i^{[M]}),\quad i=1,\ldots,N
\end{equation}
Summing over the elements of the multiplex, we obtain $\sum_i
s_i^{[\alpha]}=2S^{[\alpha]}$, where $S^{[\alpha]}$ is the total
strength of layer $\alpha$.

We also define the weighted overlap of edge $i-j$ as
\begin{equation}
o_{ij}^{\rm w} = \sum_{\alpha} w_{ij}^{[\alpha]},
\end{equation}
and, consequently, the weighted aggregated overlapping adjacency
matrix $\mathcal O^{\rm w}=\{o_{ij}^{\rm w}\}$.  We can also compute
the weighted overlapping degree of node $i$ as
\begin{equation}
o_i^{\rm w}= \sum_{j}o_{ij}^{\rm w}=\sum_{\alpha}s_i^{[\alpha]}. 
\end{equation}
Summing over all nodes, we obtain
\begin{equation}
\sum_i o_i^{\rm w} = \sum_{\alpha} \sum_i s_i^{[\alpha]}= 2
\sum_{\alpha}S^{[\alpha]}=2O^{\rm w},
\end{equation}
where $O^{\rm w}$ is the size of the weighted overlapping network,
i.e. the total number of edges in the multiplex.

We would like to stress once more that the
  superposition of different layers with links of different types
  confers a non-negligible added value to multi-layered systems, which
  is lost by considering exclusively the aggregated matrices $\mathcal
  A$ and $\mathcal O$. As we will show in in Sections IV-VIII, a
  proper description of basic multiplex quantities such as degree,
  node clustering and reachability, cannot disregard the explicit or
  implicit presence of the layer index $\alpha$, and of vectorial
  variables like $\bm{k_i}$ and $\bm{A}$.

%DATASET
\section{\textbf{THE MULTI-LAYER NETWORK OF INDONESIAN TERRORISTS}}
\label{section:indonesian}
As a case of study, in this work we focus on the multiple relations
among Indonesian terrorists belonging to the so-called Noordin Top
Terrorist Network \cite{Noordin}. This data set includes information
about trust (T), operational (O), communication (C) ties and business
(B) relations among a group of 78 terrorists from Indonesia active in
recent years. In this data set, information for some of the layers can
be split into a deeper level. This is the case of the trust and
operational networks which are composed by four sub-layers each,
making them multiplexes inside a multiplex.  Layer T is obtained as
superposition of classmates, friendship, kinship and soul-mates ties,
while layer O can be split into logistic, meetings, operations and
training sub-layers.  As a first approach we represent this system as
a multiplex network with $M=4$ layers, namely T, O, C and B.
\begin{table}[h] 
\begin{center} 
  \begin{tabular}{|c|c|lc|ccc|} 
\hline
   LAYER & CODE & $N_{act}$ & $K$ & $S$ & $O$ & $O^{\rm w}$ \\
   \hline
   \hline
   MULTIPLEX  & M & 78  & 623 &/ & 911 & 1014 \\
   \hline
   \hline
   TRUST & T &  70 & 259 &293 & / & /\\
   \hline
   Classmates & Tc &  39 & 175&/ & / & / \\
   \hline
   Friendship& Tf &  61 & 91&/ & / & /\\
   \hline
   Kinship & Tk & 24 & 16&/ & / & /\\
   \hline
   Soulmates & Ts & 9 & 11&/ & / & /\\ 
   \hline
   \hline
   OPERATIONAL &O &  68 & 437&506 & /  &/\\
    \hline
   Logistic & Ol& 16 & 29&/ & / & / \\
   \hline
   Meetings & Om&  26 & 63&/ & / & /\\
   \hline
   Operations & Oo & 39 & 267 &/& / & /\\
   \hline
   Training & Ot & 38 & 147 & /& / & /\\ 
   \hline
   \hline
   COMMUNICATION & C & 74 & 200&200 & / &/ \\
   \hline
   \hline
   BUSINESS & B & 13 & 15 &15& / &/ \\            
\hline
\end{tabular}
  \caption[]{The Top Noordin Terrorist Network includes data about
    trust (T), operations (O), communication (C) and business (B)
    among 78 terrorists active in recent years in Indonesia. Trust and
    operational networks are characterized by a deeper internal
    structure, and they can be divided into four sub-layers each. 
    For the multiplex network (M), and each layer and
    sub-layer we show the total number of active nodes $N_{act}$, and
    the number of edges expressed as non-overlapping links $K$,
    overlapping links $O$ and weighted overlapping links $O^{\rm
      w}$. For each layer $\alpha$ we also report the total strength
    $S^{[\alpha]}$.}
\label{tabledata1}
\end{center}
\end{table}
We exploit the additional richness of the data set to assign a weight
to the links connecting nodes in layers T and O, while we leave the
analysis of multiplexes of multiplexes for future work. In particular,
we associated an integer number $w_{ij}^{[\rm T]}$ with $1\le
w_{ij}^{[T]}\le 4$ to every edge in the trust layer, based on how many
times the connection appears in the four corresponding
sub-layers. Analogously, an integer weight $w_{ij}^{[\rm O]}$ with
$1\le w_{ij}^{[\rm O]}\le 4$ is associated to every edge in the
operational layer. For most of the following analysis we will consider
also T and O as unweighted layers, while we will make explicit use of
the weights of layers T and O in Section~\ref{section:overlap}.

\begin{figure}[!t]
  \center
  \includegraphics[width=3in]{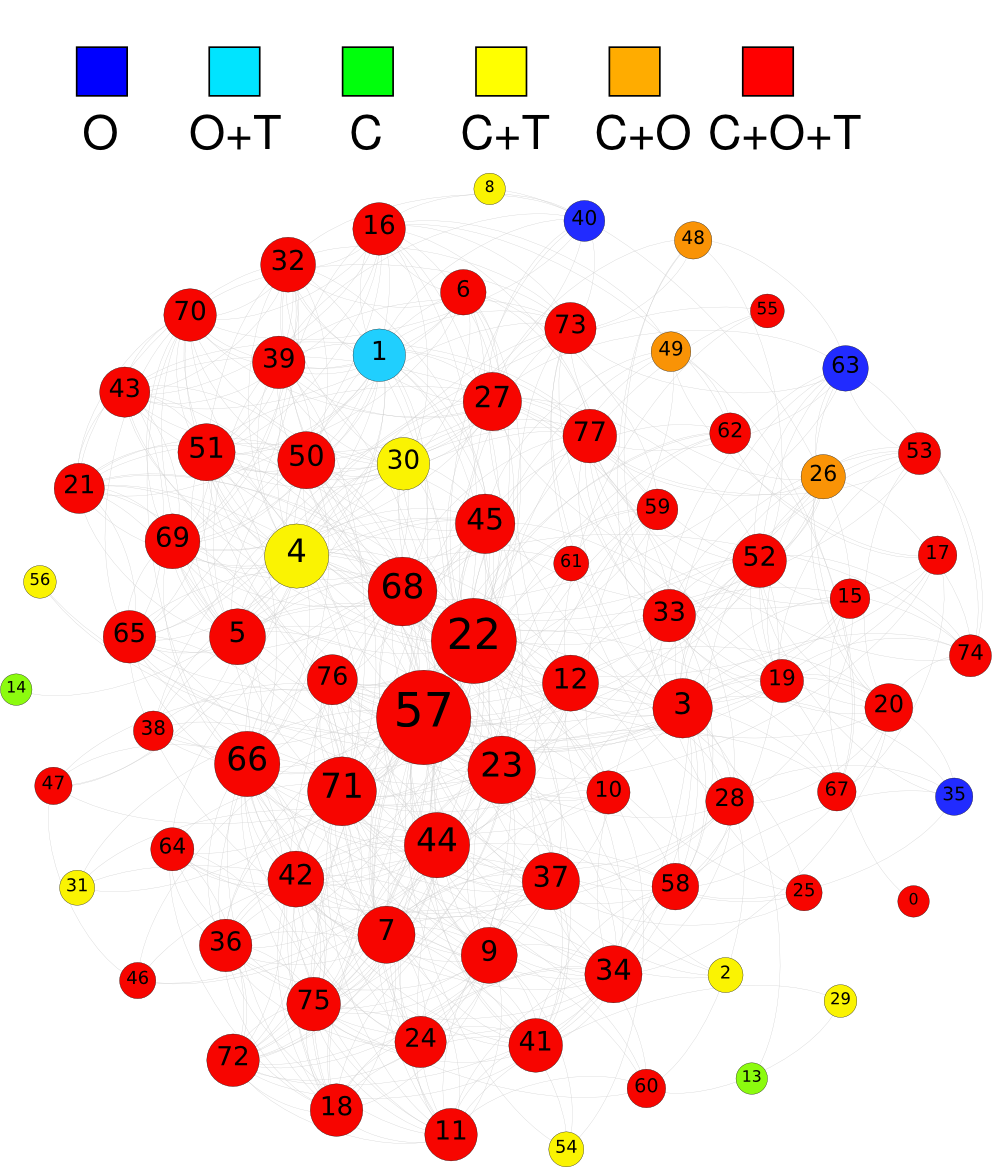}
  \caption{(color online) A flattened representation of the
    three-layer multiplex obtained by considering only trust (T),
    communication (C) and operations ties (O). For each node $i$ we
    indicate with a color-code the layers in which $i$ is actively
    involved (i.e., the layers $\alpha$ for which
    $k\lay{\alpha}_i>0$). The size of a node is proportional to its
    overlapping degree $o_i$: node $57$ is the node with the largest
    overlapping degree.}
  \label{fig:figure1}
\end{figure}

Summing up, the multiplex network of Top Noordin Terrorists has $N=78$
nodes, $K=623$ non-overlapping links, $O=911$ overlapping links and
$O^{\rm w}=1014$ weighted overlapping links.  Table~\ref{tabledata1}
reports more details about the size of each layer and sub-layer, and
of the corresponding aggregated adjacency matrices.
We notice that some individuals are not involved in
  all the four layers, meaning that their activity with respect to a
  particular kind of social relationship has not been registered or
  was unknown at the time the data set was compiled. Consequently,
  some of the replicas of such nodes will be isolated nodes on one or
  more of the four layers.  It is evident that while the trust,
communication and operational layers share approximately $90\%$ of the
nodes, the business layer has only 13 active nodes. Consequently, in
the following we will consider only trust, communication and
operational relationships, with the exception of
Section~\ref{section:overlap} where we will also briefly discuss the
role of the business layer. For this three-layer multiplex network we
have $N=78$, $K=620$, $O=896$ and $O^{\rm w}=999$. A schematic
(aggregated) representation of this multiplex network is reported in
Fig.~\ref{fig:figure1}. The node color-code indicates the layers in
which nodes are involved, while the size of each node is proportional
to its overlapping degree $o_i$. Notice that most of the nodes
participate to all the three layers, while just a few of them are
present in only one or two layers.

%NODE PROPERTIES
\section{\textbf{BASIC NODE PROPERTIES}}
\label{section:node}
One of the simplest features of a single-layer network is its degree
distribution. For multiplex networks, we can study how the degree is
distributed among the different nodes at each layer, but it is also
important to evaluate how the degree of a node is distributed across
different layers.
\begin{figure*}[!htbp]
\begin{center}
\includegraphics[width=6in]{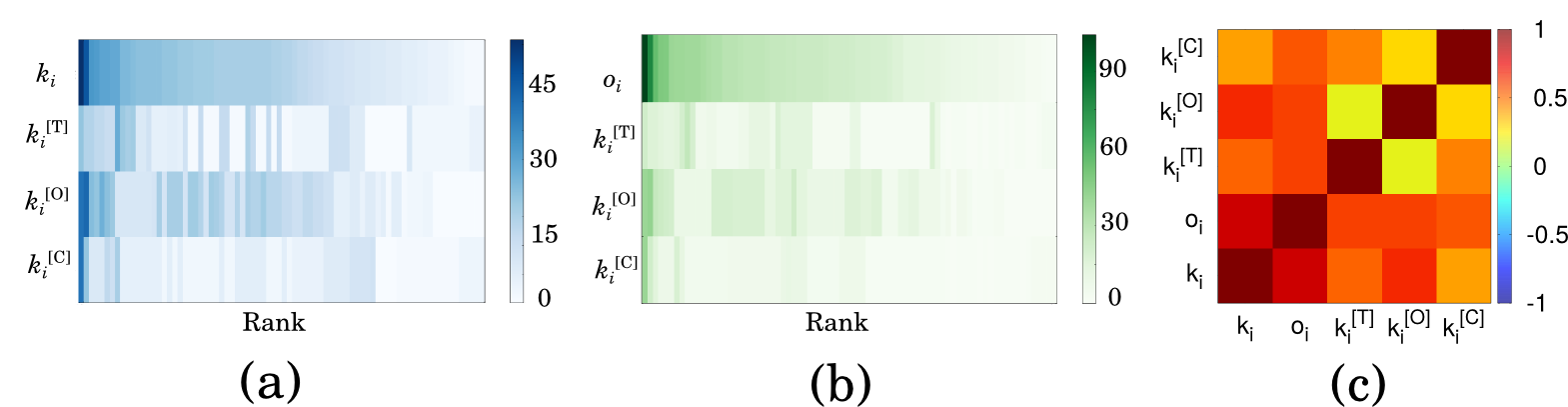}
\caption[]{(color online) (a) The top row of the panel shows with a
  color-code the degree $k_i$ of each node, from the largest (darkest,
  leftmost) to the smallest (brightest, rightmost). Keeping fixed the
  ranking induced by $k_i$, in the other three rows we report
  respectively the degree in the trust layer $k_i^{[\rm T]}$,
  operational layer $k_i^{[\rm O]}$ and communication layer $k_i^{[\rm
      C]}$. (b) Same as panel (a) but in the first row nodes are
  ranked according to their overlapping degree $o_i$. (c) The heat map
  represents the Kendall $\tau$ correlation coefficient between $k_i$,
  $o_i$, $k_i^{[\rm T]}$, $k_i^{[\rm O]}$ and $k_i^{[\rm C]}$. Notice
  that the degree of a node in the operational layer O is poorly
  correlated with its degree in the communication and trust layers
  (bright yellow regions in the heat map).}
\label{fig:figure2}
\end{center}
\end{figure*}
It is in fact possible that nodes which are hubs in one layer have
only few connections, or are even isolated, in another layer. Or,
alternatively, nodes which are hubs in one layer are also hubs in the
other layers.  We have therefore computed the aggregated topological
degree $k_i$ and the degree of the nodes in each layer 
$k_i^{[\alpha]}$, with $\alpha\in\{\rm T, O, C\}$, ranking the nodes
according to their aggregated topological degree. In
Fig.~\ref{fig:figure2}(a) we compare with a color-code plot the values
of $k_i$ with the values $k_i^{[\alpha]}$ of the node degree at each
layer $\alpha$. By visual inspection, the four degree sequences appear
weakly correlated, with nodes which are hubs in one level often having
only few connections in another layer.
In Fig.~\ref{fig:figure2}(b) we report the results obtained by ranking
the nodes according to their overlapping degree $o_i$. Also in this
case we observe weak correlations between the four degree
sequences. To better quantify such correlations, we computed the
Kendall rank correlation coefficient, $\tau_{k}$, which measures the
similarity of two ranked sequences of data $X$ and $Y$. The
correlation coefficient $\tau_{k}$ is a non-parametric measure of
statistically dependence between two rankings, since it does not make
any assumption about the distributions of $X$ and $Y$, and takes
values in $[-1,1]$. We get $\tau_{k}(X,Y)=1$ if the two rankings are
identical, $\tau_{k}(X,Y)=-1$ if one ranking is exactly the reverse of
the other and finally $\tau_{k}(X,Y)=0$ if $X$ and $Y$ are
independent. In Fig.~\ref{fig:figure2}(c) we report as a heat map the
values of $\tau_{k}$ obtained for the rankings of each pair of
variables. Notice that the aggregated degrees $k_i$ and $o_i$ are
usually weakly correlated with the degree of node $i$ on each single
layer. The highest correlation is indeed found between the degree of
the aggregated topological network $k_i$ and the overlapping degree
$o_i$.

Due to the heterogeneity in the degree distribution of each layer and
to the weak correlation observed between the degrees of the same node
at different layers, it is necessary to introduce a measure to
quantify the richness of the connectivity patterns across layers.  For
instance, consider two nodes $i$ and $j$ having exactly the same value
of overlapping degree $o_i = o_j$, and imagine that $i$ is a massive
hub on a layer $\alpha$ and an isolated node on the other layers, so
that $o_i = k\lay{\alpha}_i$, while $j$ has the same number of edges
on each layer, so that $o_j=M k\lay{\alpha}_j, \forall \alpha$. From a
multiplex perspective $i$ and $j$ have radically different roles, but
this fact is not detectable by comparing their overlapping degrees,
which have the same value. Conversely, even if $o_i$ and $o_j$ are
very different, $i$ and $j$ can look very similar if one considers the
contribution of each layer to the total overlapping degree of the two 
nodes.  

A suitable quantity to describe the distribution of the degree of node
$i$ among the various layers is the entropy of the multiplex degree:
\begin{equation}
H_i
=-\sum_{\alpha=1}^M\frac{k_i^{[\alpha]}}{o_i}ln\biggl(\frac{k_i^{[\alpha]}}{o_i}\biggr).
\label{nodeentropy}
\end{equation}
This entropy is equal to zero if all the links of node $i$ are in a
single layer, while it takes its maximum value when the links are
uniformly distributed over the different layers. In general, the
higher the value of $H_i$, the more uniformly the links of node $i$
are distributed across the layers. A similar quantity is the multiplex
participation coefficient $P_i$ of node $i$:
\begin{equation}
P_i=\frac{M}{M-1}\left[1-
  \sum_{\alpha=1}^M\biggl(\frac{k_i^{[\alpha]}}{o_i}\biggr)^2\right].
\label{participationcoefficient}
\end{equation}
\begin{figure*}[!t]
\begin{center} 
\includegraphics[width=6in]{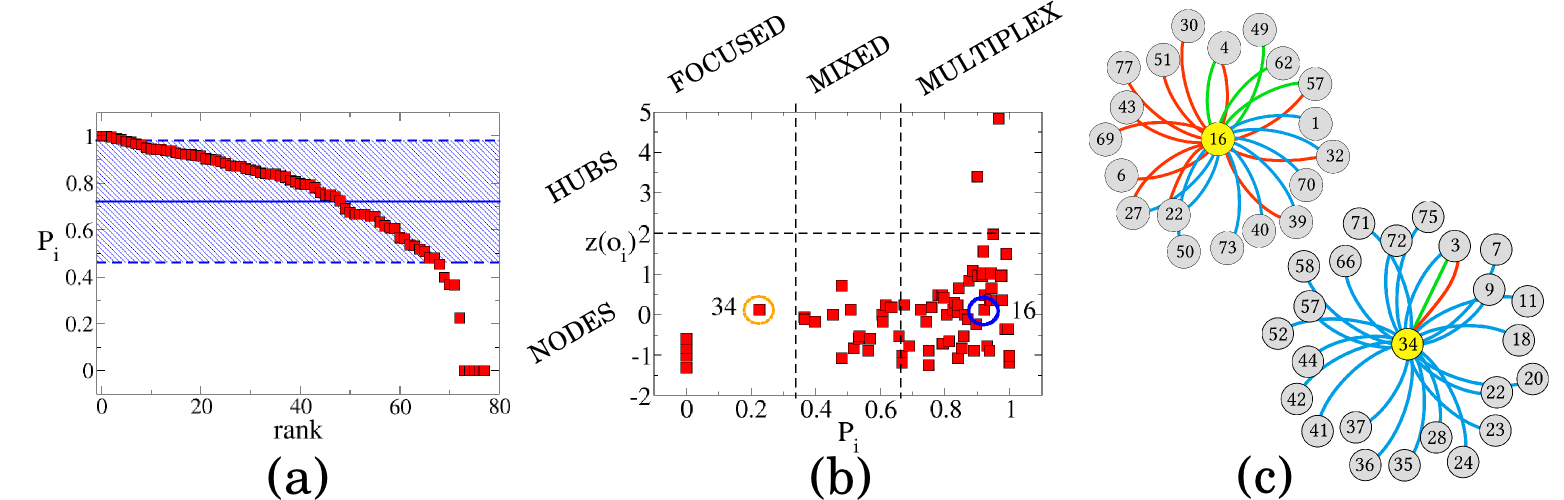}
\caption[]{(color online) (a) Rank distribution of the participation
  coefficient $P_i$ for the multi-layer network of Top Noordin
  Indonesian Terrorists. $M=3$ layers were considered, namely trust,
  operational and communication. The average value $P=0.72$ is shown
  as a horizontal blue line, while the shaded band indicates the
  standard deviation. (b) A cartography of the roles of the nodes in a
  multi-layer network can be obtained by plotting, for each node $i$,
  the multiplex participation coefficient $P_i$ versus the Z-score of
  the total overlapping degree $z(o_i)$. Even if two nodes have
  exactly the same value of $z(o_i)$ (like node $16$ and node $34$,
  indicated by the orange and blue circles, respectively), they can
  have pretty different roles, according to the value of the multiplex
  participation coefficient. (c) The ego networks of node $16$ and
  $34$, in which edges are colored according to the layer to which
  they belong, respectively green (trust), blue (operational) and red
  (communication). It is evident that the connectivity pattern of node
  $16$, whose links are homogeneously distributed across the three
  layers, is ``more \textit{multiplex}'' than that of node $34$, which
  is instead \textit{focused} on the operational layer.}
\label{fig:figure3}
\end{center} 
\end{figure*}
%%%%%%%%%%%%%%%%%%%%%

The definition of the multiplex participation coefficient is in the
same spirit of that of participation coefficient introduced in
Refs.~\cite{cartography, metaboliccartography} to quantify the
participation of a node to the different communities of a network. In
this adaptation to multi-layer networks, $P_i$ takes values in $[0,1]$
and measures whether the links of node $i$ are uniformly distributed
among the $M$ layers, or are instead primarily concentrated in just
one or a few layers. Namely, the coefficient $P_i$ is equal to $0$
when all the edges of $i$ lie in one layer, while $P_i=1$ only when
node $i$ has exactly the same number of edges on each of the $M$
layers. In general, the larger the value of $P_i$, the more equally
distributed is the participation of node $i$ to the $M$ layers of the
multiplex. The participation coefficient $P$ of the whole multiplex is
defined as the average of $P_i$ over all nodes, i.e. $P=1/N\sum_i
P_i$. The two quantities $P_i$ and $H_i$ give very similar
information, so that in the following we will discuss the results for
$P_i$ only.

In Fig.~\ref{fig:figure3}(a) we plot the distribution of $P_i$ for the
multi-layer network of Indonesian terrorists under study. Although the
average participation coefficient of the multiplex is equal to
$P=0.72$, we observe a quite broad distribution of $P_i$ in the range
$[0, 1]$.  This variance suggests the existence in the network of
various levels of node participation to each of the three layers.
Since the overlapping degree of a node represents its overall
importance in terms of number of incident edges, while the multiplex
participation coefficient gives information about the distribution of
incident edges across the layers, we propose to classify the nodes of
a multiplex by looking, at the same time, at their multiplex
participation coefficient and at their overlapping degree. With
respect to the multiplex participation coefficient, we identify three
classes. We call \textit{focused} those nodes for which $0\le P_i\le
1/3$, \textit{mixed} the nodes having $1/3<P_i\le 2/3$ and
\textit{truly multiplex} (or even simply \textit{multiplex}) the nodes
for which $P_i > 2/3$. Instead of the overlapping degree we consider
the associated Z-score, which allows to compare multiplex networks of
different size:
\begin{equation}
z(o_i)=\frac{o_i- \langle o \rangle}{\sigma_o}
\end{equation}
where $\langle o \rangle$ is equal to the average overlapping degree
of the nodes of the system, and $\sigma_o$ is the corresponding
standard deviation. With respect to the Z-score of their overlapping
degree, we distinguish \textit{hubs}, for which $z(o_i)\ge 2$, from
regular nodes, for which $z(o_i)<2$. Consequently, by considering the
multiplex participation coefficient $P_i$ of a node and its total
overlapping degree $o_i$ we can define six classes of nodes, as
depicted in Fig.~\ref{fig:figure3}(b), where we represent each node as
a point in the $(P_i, z(o_i))$ plane.  
\begin{figure*}[!t]
\begin{center}
\includegraphics[width=6in]{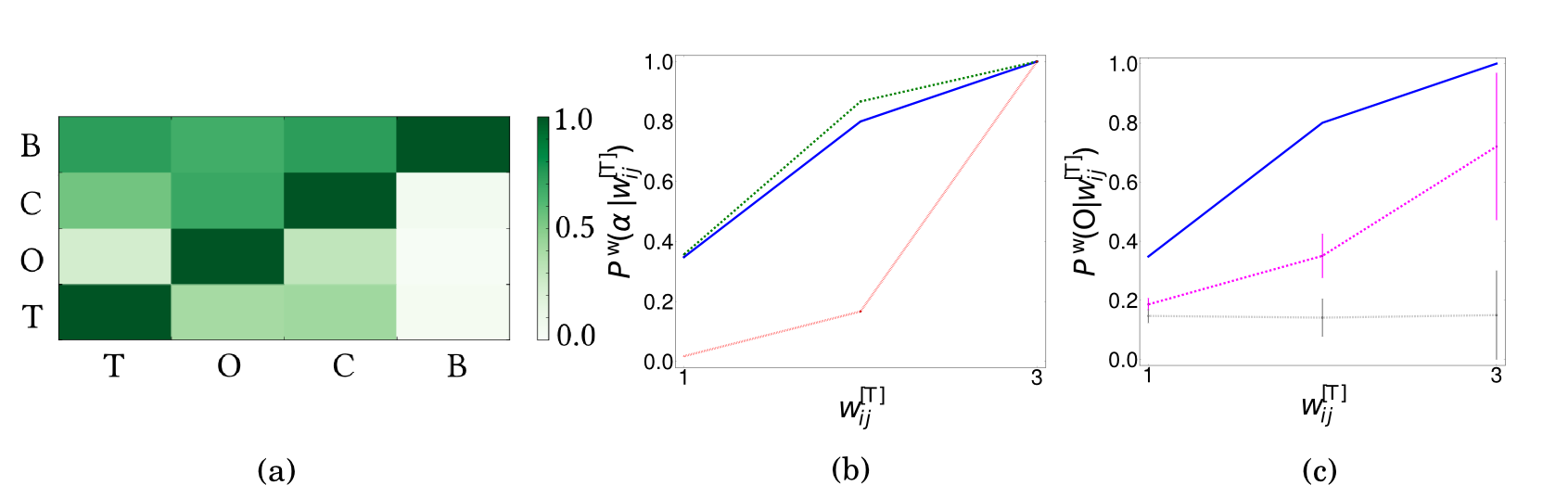}
\caption{(color online) (a) For each layer $\alpha$, we show in the
  color map the fraction of edges which is also present in each other
  layer $\alpha'$. (b) Probability $P^{\rm w}$ of finding a certain
  link at layer O (solid blue line), C (dotted green line)and B (dashed red line), conditional to the weight $w_{ij}^{[\rm
      T]}$ of the same link at layer T.  (c) The
    values of $P^{\rm w}$ computed on real data for
    layer O (solid blue line) are compared to those obtained by randomizing the edges
    keeping fixed the total number of links $K^{[\rm O]}$ (dashed grey line) or the
    degree distribution $P(k^{[\rm O]})$ (dotted magenta line) for the operational layer.}
\label{fig:figure4}
\end{center}
\end{figure*}
Notice that the distribution of $z(o_i)$ is asymmetric and unbalanced
towards positive values, and this is a sign of the heterogeneity of
the total overlapping degree. Moreover, there is a quite large
heterogeneity in the values of $P_i$ for a fixed value of
$z(o_i)$. Let us focus for instance on two specific nodes, namely node
$16$ and $34$. These two nodes have the same overlapping degree,
namely $o_{16}=o_{34}= 25$, corresponding to
$z(o_{16})=z(o_{34})=0.12$~, but very different participation
coefficient across layers T, O and C, respectively $P_{16}=0.915$ and
$P_{34}=0.23$. Consequently, even if the overall number of edges of
node $16$ and node $34$ is the same (which would make these two nodes
indistinguishable in the aggregated overlapping network), they play
radically different roles, as becomes evident by looking at their ego
networks, reported in Fig.~\ref{fig:figure3}(c). In fact, while node
$34$ is highly focused on the operational layer (blue edges), with
only one edge in the trust layer (green edge) and one edge in the
communication layer (red edge), node $16$ is instead involved in all
the three layers, with a comparable number of edges in each of
them. This implies that the removal of node $34$ would primarily
affect just the operational layer, while the absence of node $16$
could cause major disruptions in the trust, operational and
communication networks.  Similar results are obtained by considering
the Z-score of the degree $k_i$ of node $i$ in the aggregated
topological network (figure not shown).

\section{\textbf{EDGE OVERLAP AND SOCIAL REINFORCEMENT}}
\label{section:overlap}

After having proposed some measures of the role of individual nodes in
the multiplex, we now aim at quantifying the importance of each layer
as a whole.  For instance, we can detect the existence of correlations
across the layers of a multiplex by computing the edge overlap
$o_{ij}$ of Eq.~(\ref{edge_overlap}) for each edge $i-j$, and by
looking at how this quantity is distributed.  We now consider the
multiplex formed by all the four layers of the Noordin Indonesian
Terrorist Network, i.e. the trust, operational, communication and
business layers, so that $1\le o_{ij}\le 4$ for all possible pairs of
nodes connected by at least one edge. If we look at the distribution
of $o_{ij}$, we see that $46\%$ of the edges exist in just one of the
four layers, $27\%$ are present in two layers, $23\%$ exist in three
layers and only $4\%$ are present in all the four layers.

Besides the distribution of $o_{ij}$ gives some information about the
existence of inter-layer correlations, it is not able to disentangle
the relevance of single layers.
A slightly more sophisticated quantity we can look at is the
conditional probability of finding a link at layer $\alpha'$ given the
presence of an edge between the same nodes at layer $\alpha$:
\begin{equation}
P(a_{ij}^{[\alpha']}|a_{ij}^{[\alpha]})=\frac{\sum_{ij}a_{ij}^{[\alpha']}a_{ij}^{[\alpha]}}{\sum_{ij}a_{ij}^{[\alpha]}}
\label{eq:prob_cond}
\end{equation}
The denominator of Eq.~(\ref{eq:prob_cond}) is equal to the number
$K^{[\alpha]}$ of edges at layer $\alpha$, while the numerator is
equal to the number of such edges which are also present at the layer
$\alpha'$.  The conditional probability
${P}(a_{ij}^{[\alpha']}|a_{ij}^{[\alpha]})$ is shown as
a heat-map in Fig.~\ref{fig:figure4}(a) for the four layers.
For instance, the first column shows with a color-code the probability
to find a link on layer T given its existence on layer B, C, O or T
(obviously, we have ${P}(a_{ij}^{[T]}|a_{ij}^{[T]})=1$), while the
last row represents the fraction of edges in layer T which also exist
in layer T, O, C and B. Since layers T and O have a composite internal
structure of four levels each, which allows us to assign a weight
$w_{ij}^{[\rm T]}$ and $w_{ij}^{[\rm O]}$ to each pair of connected
nodes $i$ and $j$, it is interesting to study the probability ${P^{\rm
    w}}(a_{ij}^{[\alpha']}|w_{ij}^{[\alpha]})$ of having a link on
layer $\alpha'$ given its weight on the leading layer $\alpha$, with
$\alpha$ corresponding to layers O, and T. In
Fig.~\ref{fig:figure4}(b) we plot the probability of finding a link at
layer O, C and B, given the weight $w_{ij}^{[\rm T]}$ of the link at
layer T.  Even though in principle $w_{ij}^{[\rm T]}=4$ is possible,
none of the edges appears together in all classmates, friendship,
kinship and soul-mates sub-layers of the trust layer. In all the three
cases, ${P^{\rm w}}$ is an increasing function of
$w_{ij}^{[\rm T]}$. Fig.~\ref{fig:figure4}(b) suggests that the
stronger the trust connection between two terrorists the higher the
probability for them to operate together, communicate or having common
business. In particular, for layer O and C, which are the ones that
have a number of nodes comparable to the one of layer T, already a
value of $w_{ij}^{[\rm T]}=2$ implies that the two people have common
operations and communications in $80\%$ of the cases.  If
$w_{ij}^{[\rm T]}=3$, then the probability that the edge $i-j$ exists
in all the three remaining layer is equal to $1$.
\begin{figure*}[!t]
\begin{center}
\includegraphics[width=6in]{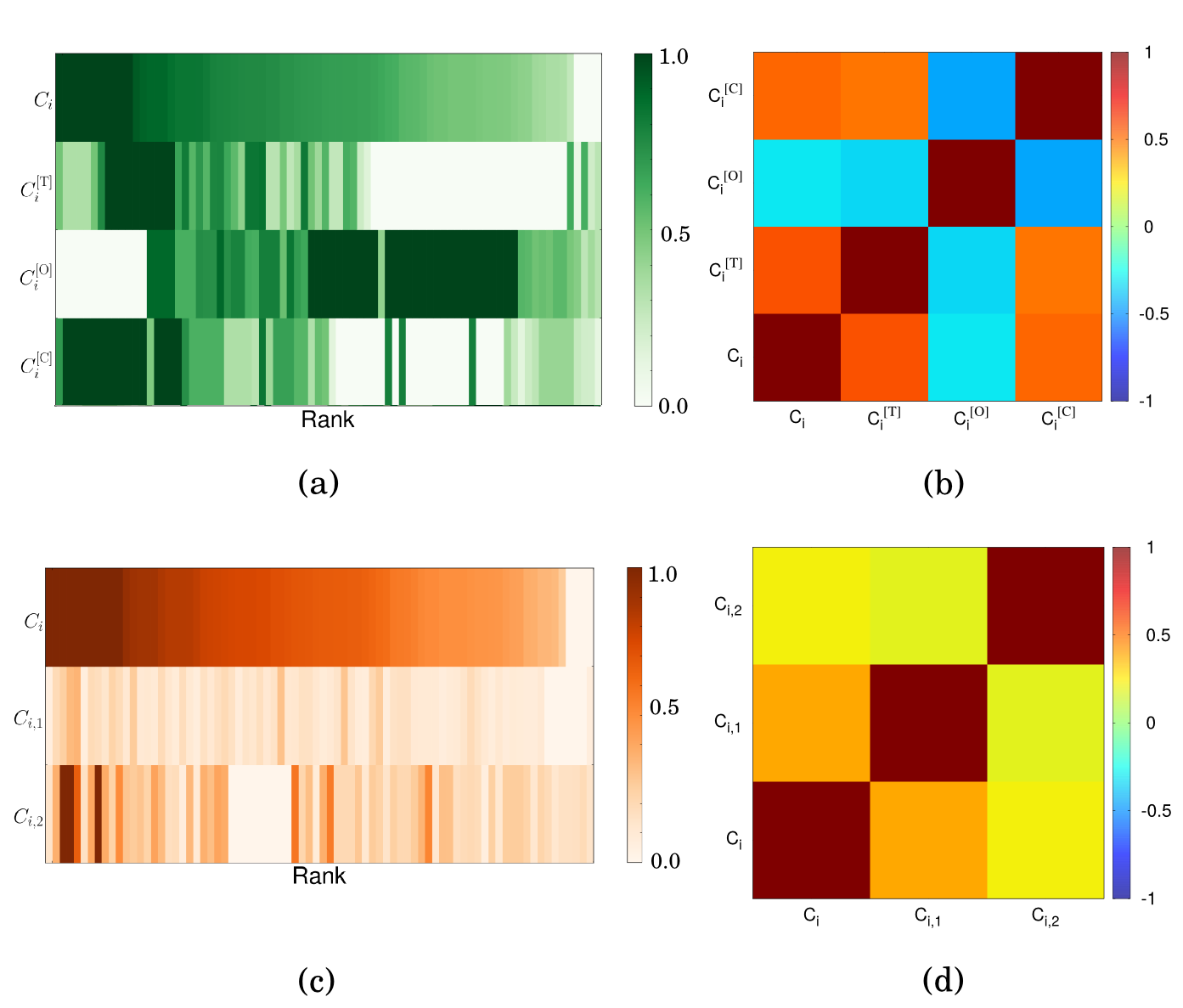}
\caption[]{(color online) (a) The node clustering coefficient $C_i$ of
  the aggregated topological network and of the three layers T, O, C,
  respectively denoted as $C_i^{[\rm T]}$, $C_i^{[\rm O]}$ and
  $C_i^{[\rm C]}$. The nodes are ranked according to their value of
  $C_i$ on the aggregated topological network. (b) The heat map
  represents the correlation between the rankings of nodes according
  to their clustering coefficients on the three layers and on the
  topological aggregated network. Notice that $C_i^{[\alpha]}$ is
  weakly correlated with $C_i$ for $\alpha\in\{T,O,C\}$, and that such
  correlation might also be negative, as in the case of $C_i^{[\rm
      O]}$. (c) Comparison among the clustering coefficient
  $C_i$ of the aggregated topological network, and the multi-layer
  clustering coefficients $C_{1,i}$ and $C_{2,i}$. The nodes are ranked 
  according to their value of $C_i$. (d) The heat map represents the 
  correlation between the rankings of nodes according to $C_i$, 
  $C_{1,i}$ and $C_{2,i}$.}
\label{fig:figure5}
\end{center}
\end{figure*}
%
%%%%%%%%%%%%%%%%%%%

This phenomenon can be explained in terms of social reinforcement,
meaning that the existence of strong connections in the Trust layer,
which represents the strongest relationships between two people,
actually fosters the creation of links in other layers and produces a
measurable effect on the probability to operate, communicate and do
business together. Despite we do not have longitudinal information to
test the hypothesis that original trust connections actually caused
the creation of links in other layers by means of social
reinforcement, in this particular case we have to stress that the
strength of the trust relationship between two individuals is higher
if they had been kin, classmates, soul-mates, and/or friends,
respectively~\cite{Noordin}. This means that, with high probability,
the establishment of any of the four Trust relationships between $i$
and $j$ preceded by several years the establishment of any
communication, operational or business relationship registered during
the collection of the data set. Consequently, it is not too
pretentious to suggest that a social reinforcement mechanism took
place in this small social system, and that trust relationships have
actually \textit{caused} the subsequent communication and the
collaboration among the terrorists.
\begin{figure}[!t]
\begin{center}
\includegraphics[width=3in]{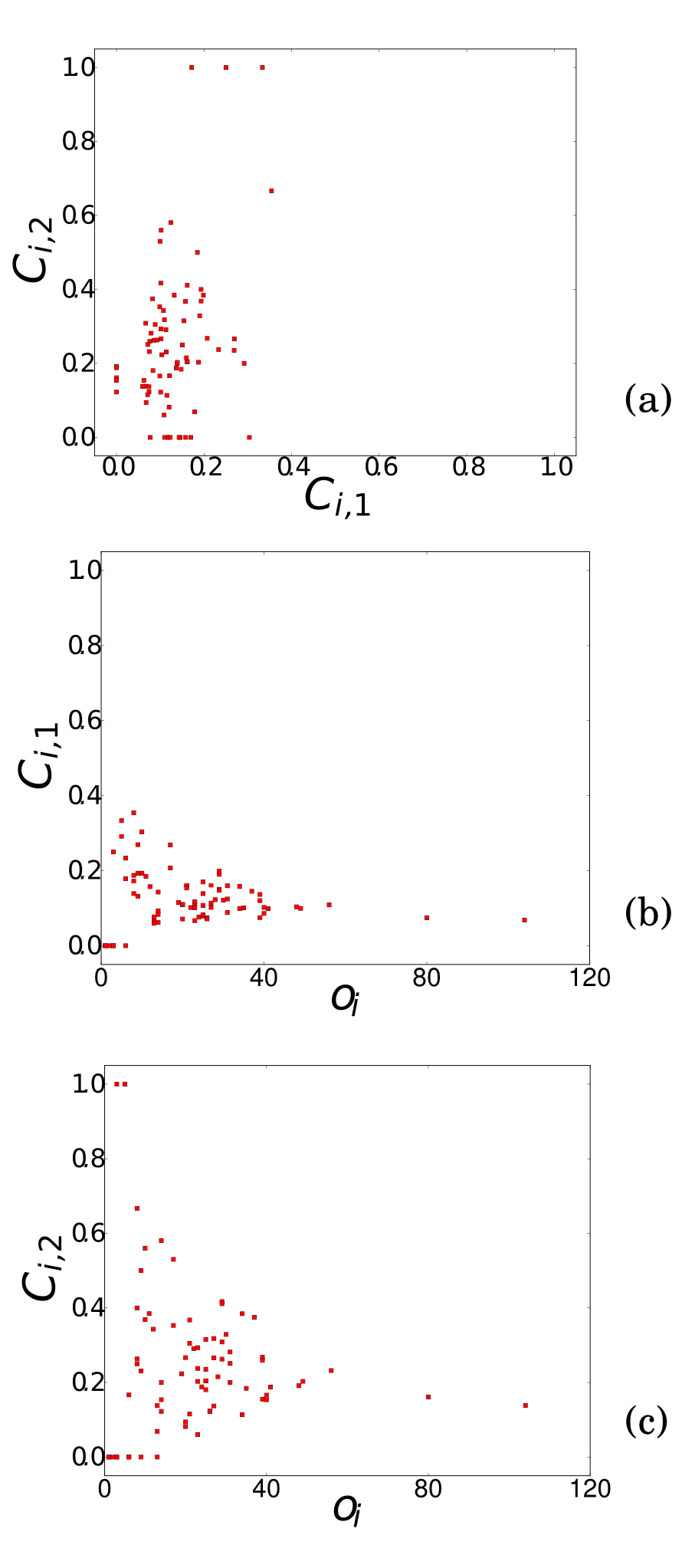}
\caption[]{(color online) Scatter-plots of (a) $C_{1,i}$ versus
  $C_{2,i}$ (b) $C_{1,i}$ versus $o_i$ and (c) $C_{2,i}$ versus
  $o_i$. The values of the Kendall's $\tau$ and of the Pearson's
  linear correlation coefficient $r$ for any pair of measures are,
  respectively: ${\tau(C_{i,1},C_{i,2})=0.61}$,
  ${r(C_{i,1},C_{i,2})=0.76}$,
  ${\tau(C_{i,1},o_i)=-0.11}$,
  ${r(C_{i,1},o_i)=-0.13}$,
  ${\tau(C_{i,2},o_i)=0.01}$,
  ${r(C_{i,2},o_i)=0.04}$. It is worth noticing that
  both $C_{i,1}$ and $C_{2,i}$ are almost uncorrelated with the
  overlapping degree $o_i$, a fact that confirms their truly multiplex
  nature.}
\label{fig:figure6}
\end{center}
\end{figure}
%
%%%%%%%%%%%%%%%

In order to statistically validate these results, in
  Fig.~\ref{fig:figure4}(c) we report the expected values of $P^{\rm
    w}$ obtained by randomizing the non-leading layers while keeping
  fixed either the total number of links $k^{[\alpha]}$ or the degree
  distribution $P(k^{[\alpha]})$. In the first case, each non-leading
  layer is an \Erdos-Renyi random graph and $P^{\rm w}$ is not even
  correlated with the weights on layer T, as expected. In the second
  case, which is an extension to multiplexes of the configuration
  model, for each weight $w_{ij}^{[\rm T]}$ the conditional
  probability to find an edge on the operational layer is
  systematically lower in the randomized networks than in the original
  one. Hence, we can conclude that inter-layer correlations among the
  heterogeneous degree distributions of the various levels do not
  provide an ultimate explanations to the founded results for $P^{\rm
    w}$ and that the Trust layer is genuinely \textit{driving} the
  observed connection pattern. Results analogous to that of layer O,
were also found for layers C and B.

Similar results are obtained considering the operational network
(instead of the trust network) as leading layer, but in this case the
conditional probability of finding an edge in T, C and B given its
weight in O was substantially smaller than those reported in
Fig.~\ref{fig:figure4}(b) and Fig.~\ref{fig:figure4})(c)(figure not
shown). This is not surprising at all, since while it is clear that a
stronger level of trust between two individuals can boost their
communications and their common operations, we expect a weaker
causality between the strength of different operations two individuals
have shared and their trust and communications. The existence of a
weaker interaction between the operational layer and the other three
layers increases the validity of our hypothesis that the trust layer
is indeed controlling the overall structure of the multiplex network
through a social reinforcement mechanism and that the
  relative importance of the trust layer for the formation of edges on
  other layers is not a mere consequence of existence of sub-layers.

\section{\textbf{TRANSITIVITY AND CLUSTERING}}
\label{section:clustering}
One of the most remarkable characteristic of complex real-world
single-layer networks, especially acquaintance and collaboration
networks, is the tendency of nodes to form triangles, i.e. simple
cycles involving three nodes. This widely observed tendency is
concisely expressed by the popular saying \textit{``the friend of your
  friend is my friend''} and is usually quantified through the
so-called node clustering coefficient~\cite{clustering}. The
clustering coefficient of node $i$ is defined as :
\begin{equation}
C_i = \frac{\sum_{j \neq i, m \neq i} a_{ij}a_{jm}a_{mi}}{\sum_{j \neq
    i, m \neq i} a_{ij}a_{mi}}=\frac{\sum_{j \neq i, m \neq i}
  a_{ij}a_{jm}a_{mi}}{k_i(k_i-1)}.
\label{Ci}
\end{equation}
and quantifies how likely it is that two neighbors of node $i$ are
connected to each other. In fact, Eq.~(\ref{Ci}) measures the fraction
of triads centered in $i$ that close into triangles. By definition
$C_i$ takes values in the interval $[0,1]$.  Averaging this quantity
over all the nodes in a network, one gets the network clustering
coefficient:
\begin{equation}
 C=\frac{1}{N} \sum_i C_i.  
\end{equation}
A similar ---although not identical--- measure of local
cohesion~\cite{latora}, which is commonly used in the social sciences,
is the network transitivity \cite{wasserman94}:
\begin{equation}
T = \frac{3 \times \text{No. of triangles in the graph}}{\text{No. of triads in the graph}}.
\end{equation}
This is defined as the proportion of triads, i.e. connected triples of nodes, 
which close into triangles. 

Since each layer of a multiplex can be seen as a single-layer network,
the definitions of network clustering coefficient and network
transitivity can be used to characterize the abundance of triangles on
each layer. In general, different layers may show similar or
dissimilar patterns of clustering. In Table~\ref{tableclustering1} we
report the average clustering coefficient and the transitivity for
each layer of the terrorist network, and for its topological
aggregate.
\begin{table}[h] 
\begin{center} 
\begin{tabular}{|c|c|c|} 
\hline
   Layer & $C$  & $T$ \\
   \hline
   T  & 0.38 & 0.53  \\
   \hline
   O &  0.67 & 0.62 \\
   \hline
   C & 0.45 & 0.27 \\
   \hline
   $\mathcal A$ & 0.66 & 0.56 \\
\hline
\end{tabular}
\caption[]{The average clustering coefficient $C$ and the transitivity
  $T$ for layers T, O, C and for the aggregated topological network
  $\mathcal A$.}
\label{tableclustering1}
\end{center}
\end{table}

Notice that each layer has quite peculiar values of clustering and
transitivity, which are in turn different from those measured on the
aggregated topological network. In particular, the highest values of
clustering and transitivity are observed in the Operations layer,
probably due to the fact that terrorist missions usually involve more
than two people at the same time.  In Fig.~\ref{fig:figure5}(a) we
focus on the node clustering coefficient, we rank the nodes of the
multiplex according to the value of $C_i$ for the aggregated
topological network and we compare this value with the clustering
coefficient calculated on each layer $C_i^{[\alpha]}$.
As shown, many nodes display quite different values of the clustering
coefficient across the layers.  We have computed the Kendall
correlation coefficient $\tau_k$ between each pair of layers and
between each layer and the topological aggregate. The results are
shown in Fig.~\ref{fig:figure5}(b), as a heat map.  Notice that at the
best the sequences of clustering coefficient are weakly correlated,
when not uncorrelated or even anti-correlated. In particular, the
ranking of clustering coefficient for the Operations layer is
anti-correlated with that of the other three layers and of the
topological aggregated network.

However, comparing the sequences of $C_i$ for each layer tells us very
little about the interplay between the several levels of the system in
terms of clustering. In particular, it is interesting to study to
which extent the multiplexicity affects the formation of triangles,
i.e. how the presence of different layers can give rise to triangles
which were impossible to close at the level of single layers. For this
reason we need to extend the notion of triangle to take into account
the richness added by the presence of more than one layer.
We define a 2-triangle a triangle which is formed by an edge belonging
to one layer and two edges belonging to a second layer. Similarly, we
call a 3-triangle a triangle which is composed by three edges all
lying in different layers. In order to quantify the added value
provided by the multiplex structure in terms of clustering, we define
two parameters of clustering interdependence $I_1$ and $I_2$. $I_1$ is
the ratio between the number of triangles in the multiplex which can
be obtained only as 2-triangles, and the number of triangles in the
aggregated system. $I_2$ is the ratio between the number of triangles
in the multiplex which can be obtained only as 3-triangles and the
number of triangles in the aggregated system. Then, $I = I_1 + I_2$ is
the total fraction of triangles of the aggregated topological network
which can not be found entirely in one of the layers. For the
multi-layer network of terrorists we obtain $I_1=0.31$ and $I_2$ of
the order of $10^{-3}$, which indicates that almost no triangle is
formed exclusively by the interplay of three different layers. This
result is suggest the presence of non-trivial patterns in clustering
and triadic closure in multi-layer systems.

In this work we also aim at generalizing the notion of clustering
coefficient to multi-layer networks. Recalling the definition of
2-triangle and 3-triangle, we define a 1-triad centered at node $i$,
for instance $j-i-k$, a triad in which both edge $j-i$ and edge $i-k$
are on the same layer. We also define a 2-triad as a triad whose two
links belong to two different layers of the systems. We are now ready
to give two definitions of clustering coefficient for multiplex
networks.  Similar definitions have been recently ---and
independently--- proposed in Ref.~\cite{arenas2}.  The first
coefficient $C_{i,1}$ is defined, for each node $i$, as the ratio
between the number of 2-triangles with a vertex in $i$ and the number
of 1-triads centered in $i$. We can express this clustering
coefficient in terms of the multi-layer adjacency matrix as:
\begin{eqnarray}
\label{clustering1eq}
C_{i,1}= \frac{\sum_{\alpha} \sum_{\alpha' \neq \alpha} \sum_{j \neq i, m \neq i } 
  (a_{ij}^{[\alpha]}a_{jm}^{[\alpha']}a_{mi}^{[\alpha]})}{(M-1) \sum_{\alpha}
  \sum_{j \neq i, m \neq i } (a_{ij}^{[\alpha]}a_{mi}^{[\alpha]})}= 
\nonumber
\\
= \frac{\sum_{\alpha} \sum_{\alpha' \neq \alpha} \sum_{j \neq i, m \neq i } (a_{ij}^{[\alpha]}a_{jm}^{[\alpha']}a_{mi}^{[\alpha]})}{(M-1)\sum_{\alpha} k_i^{[\alpha]}(k_i^{[\alpha]}-1)}
\end{eqnarray}
Since each 1-triad can theoretically be closed as a 2-triangle on each of the $M$ layers of the multiplex excluding the layer to which its edges belong, in order to have a normalised coefficient we have to divide the term by $M-1$.
In addition to this, we define a second clustering coefficient for
multiplex networks as the ratio between the number of 3-triangles with
node $i$ as a vertex, and the number of 2-triads centered in $i$. In
terms of adjacency matrices, we have:
\begin{equation}
\label{clustering2eq}
C_{i,2}= \frac{\sum_{\alpha} \sum_{\alpha' \neq \alpha} \sum_{\alpha''
    \neq \alpha, \alpha'}\sum_{j \neq i, m \neq i}
  (a_{ij}^{[\alpha]}a_{jm}^{[\alpha'']}a_{mi}^{[\alpha']}) }
{(M-2)\sum_{\alpha} \sum_{\alpha' \neq \alpha}  \sum_{j \neq i, m \neq i }  
(a_{ij}^{[\alpha]}a_{mi}^{[\alpha']})}.
\end{equation}
where a normalisation coefficient $M-2$ has been added.
While $C_{i,1}$ is a suitable definition for multiplexes with $M \ge
2$, $C_{i,2}$ can only be defined for systems composed of at least
three layers. Averaging over all the nodes of the system, we obtain
the network clustering coefficients $C_1$ and $C_2$.

In Fig.~\ref{fig:figure5}(c) we rank the nodes of the terrorist
network according to their value of $C_i$ for the aggregated system,
and compare this sequence of values with the ones obtained with the
two measures of multiplex clustering, $C_{i,1}$ and $C_{i,2}$.  As
shown in the figure, $C_{i,1}$ and $C_{i,2}$ capture different effects
of multi-layer clustering. This fact is confirmed by the heat map
reported in Fig.~\ref{fig:figure5}(d), which shows with a color-code
the non-parametric correlations among $C_{i,1}$, $C_{i,2}$ and
$C_i$. Notice that, in general, the correlation between $C_i$ and both
$C_{i,1}$ and $C_{i,2}$ is pretty small.

These results indicate that multiplex clustering provides information
which are substantially different from those obtained by looking at
the clustering of the aggregated network.  In addition to this,
$C_{i,1}$ and $C_{i,2}$ are poorly correlated, as is also evident from
Fig.~\ref{fig:figure6}(a).  In practice, for a given value of
$C_{i,1}$, we have nodes with a wide range of values of $C_{i,2}$, and
vice--versa. Consequently, it is necessary to use both clustering
coefficients in order to properly quantify the abundance of triangles
in multi-layer networks.  In Fig.~\ref{fig:figure6}(b) and
Fig.~\ref{fig:figure6}(c) we report the scatter-plots of $C_{i,1}$ and
$C_{i,2}$ versus $o_i$.  Multiplex clustering coefficients are genuine
multiplex variable and appear to be not correlated with the degree of
the nodes of the system. We also found that the clustering coefficient
is not correlated with other measures of aggregated degree, such as
$k_i$ and $o_i^{\rm w}$ (figures not shown).

We can also generalize the definition of transitivity $T$ to the case
of multi-layer networks. Similarly to the case of the clustering
coefficient we propose two measures of transitivity.  We define $T_1$
as the ratio between the number of 2-triangles and $M-1$ times the number of
1-triads in the multi-layer network. Moreover, we introduce $T_2$ as
the ratio between the number of 3-triangles and $M-2$ times the number of 2-triads
in the system. 

Notice that clustering interdependences $I_1$ and $I_2$, average
multiplex clustering coefficients $C_1$ and $C_2$ and multiplex
transitivities $T_1$ and $T_2$ are all global graph variables which
give a different perspective on the multi-layer patterns of clustering
and triadic closure with respect to the clustering coefficient and the
transitivity computed for each layer of the network. We have computed
all such quantities for the multi-layer network of the Indonesian
terrorists and, as a term of comparison, we have constructed a
configuration model for multiplex networks, which will be useful to
prove the non-trivial organization of the network under study.  

In analogy with the case of a single-layer network, for a multiplex
with $M$ layers, where each node is characterized by a degree vector
$\bm{k}_i$, we call configuration model the set of multiplexes
obtained from the original system by randomizing edges and keeping
fixed the sequence of degree vectors
$\{\bm{k}_1,\bm{k}_2,\ldots,\bm{k}_N\}$, i.e. keeping fixed the degree
sequence at each layer $\alpha$.  We can now compare the values of $C$
and $T$, $C_1$ and $C_2$, $T_1$ and $T_2$, $I_1$ and $I_2$ obtained on
real data with the average values found for the multi-layer
configuration model. The comparison is shown in
Table~\ref{tableclustering10}. As expected $C$ and $T$ computed on the
aggregated topological network for real data are systematically higher
than the ones obtained on randomized data, where edge correlations are
washed out by the randomization. For the same reason, $C_1$, $C_2$,
$T_1$ and $T_2$ are higher on real data. Conversely, we obtained
higher values on randomized data for $I_1$ and $I_2$. This is not
surprising, since the measures of clustering interdependence tell us
about the fraction of triangles which can be exclusively found as
multi-triangle in the system. Since the configuration model washes out
inter-layer correlations, it is generally easier to find
multi-triangles on a randomized multiplex network rather than on a
real one where edges have a higher overlap.
\begin{table}[h] 
\begin{center} 
\begin{tabular}{|c|c|c|} 
\hline
   Variable & Real data & Randomized data  \\
   \hline
   $C$  & 0.66 & 0.46 \\
   \hline
   $T$ & 0.56 & 0.41 \\
    \hline
   $C_1$ &  0.13 & 0.08 \\
   \hline   
   $C_2$ & 0.26 & 0.18 \\
    \hline   
   $T_1$ & 0.10 & 0.07 \\
   \hline   
   $T_2$ & 0.21 & 0.16 \\
   \hline   
   $I_1$ & 0.31 & 0.60 \\
   \hline   
   $I_2$ & 0.005 & 0.047 \\
\hline
\end{tabular}
\caption[]{Values of clustering $C$ and transitivity $T$ computed on
  the aggregated topological network, and values of the introduced
  measures for clustering in multi-layer networks, namely the
  multiplex clustering $C_1$ and $C_2$, the multiplex transitivity
  $T_1$ and $T_2$, and the clustering interdependence $I_1$ and
  $I_2$. For comparison we report also the results for a randomized
  system obtained through a multi-layer configuration model.}
\label{tableclustering10}
\end{center}
\end{table}
All these results demonstrate that, as previously shown for the
overlap, also the clustering coefficient appears to be affected by the
presence of non-trivial structural properties across the different
layers of the multiplex network under study.

\section{\textbf{REACHABILITY, SHORTEST PATHS AND INTERDEPENDENCE}}
\label{section:paths}

Reachability is an important feature in networked systems. In
single-layer networks it has to do with the existence and length of
shortest paths connecting pairs of nodes.  In multi-level systems,
shortest paths may significantly differ between different layers, and
each layer and the aggregated topological networks as well. To capture
the multiplex contribution to the reachability of each unit of the
network, the so-called node interdependence has been recently
introduced in Refs.~\cite{transport, growing}.
\begin{figure}[!t]
\begin{center}
\includegraphics[width=3in]{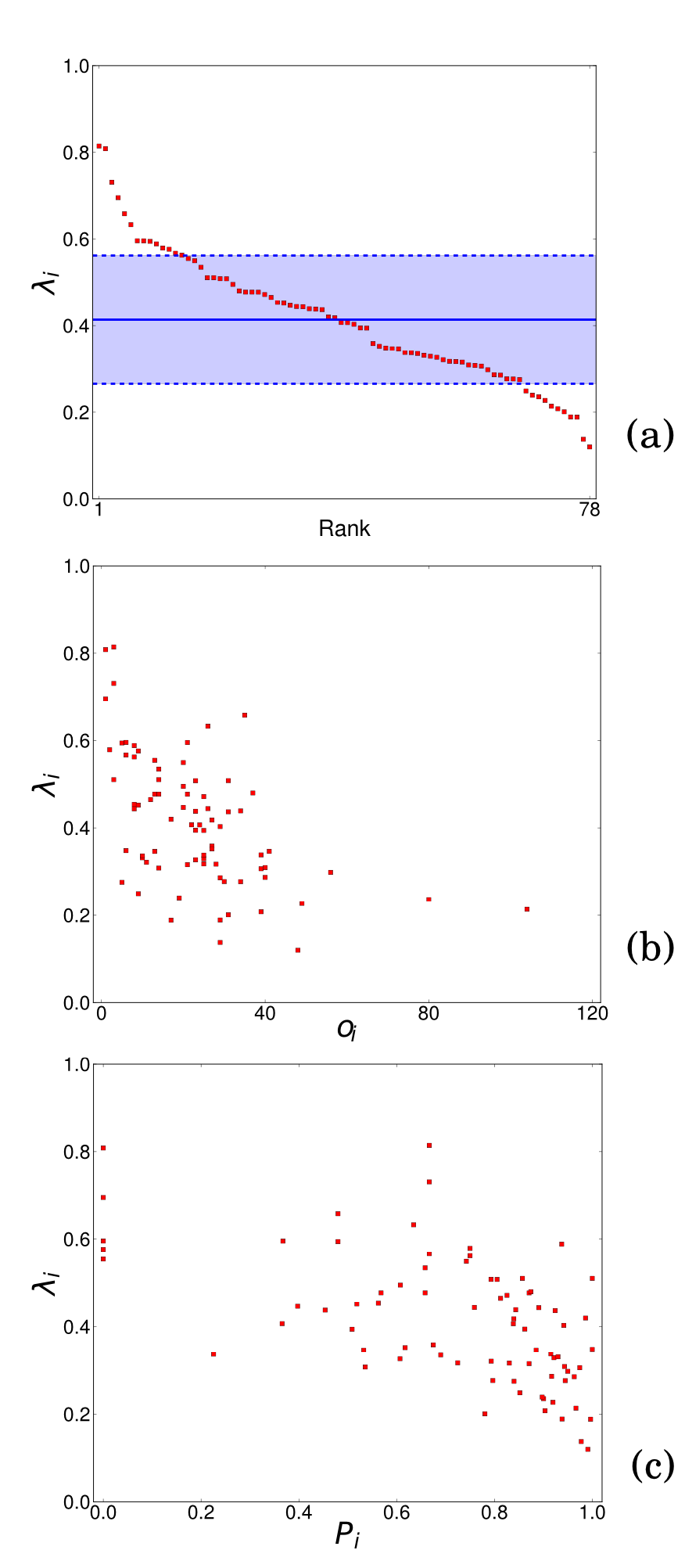}
\caption[]{ (color online) (a) Rank distribution of the node
  interdependence $\lambda_i$ in the Indonesian Terrorist multiplex
  network. (b) Scatter-plot of the interdependence $\lambda_i$ versus
  $o_i$ and (c) versus $P_i$. The corresponding value of Kendall's
  $\tau$ and Pearson's $r$ correlation coefficient are, respectively,
  ${\tau(\lambda_i,o_i)=-0.41}$, ${r(\lambda_i,o_i)=-0.56}$,
  ${\tau(\lambda_i,P_i)=-0.41}$, ${r(\lambda_i,P_i)=-0.57}$}
\label{fig:figure7}
\end{center}
\end{figure}
The interdependence $\lambda_i$ of node $i$ is defined as:
\begin{equation}
\label{interdependenceeq}
\lambda_i = \sum_{j \neq i}\frac{\psi_{ij}}{\sigma_{ij}} 
\end{equation}
where $\sigma_{ij}$ is the total number of shortest paths between node
$i$ and node $j$ on the multiplex network, and $\psi_{ij}$ is the
number of shortest paths between node $i$ and node $j$ which make use
of links in two or more than two layers.  Hence, the node
interdependence is equal to $1$ when all shortest paths make use of
edges laying at least on two layers, and equal to $0$ when each of the
shortest paths makes use of only one of the $M$ layers of the
system. Averaging $\lambda_i$ over all nodes, we obtain the network
interdependence $\lambda = (1/N) \sum_i \lambda_i$. In
Fig.~\ref{fig:figure7} we display the rank distribution of
$\lambda_i$. The network has a large variety of node
interdependencies: although most of the nodes have a value of
$\lambda_i$ in the range $[0.27, 0.56]$ around the average value
$\lambda = 0.41$, there are also nodes with values as small as
$\lambda_i=0.1$, and two nodes with values larger than $0.8$.

The interdependence is a genuine multiplex measure and, as shown in
Fig.~\ref{fig:figure7}(b) provides information in terms of
reachability which is slightly anti-correlated to measures of degree
such as $o_i$. In fact, a node with a high overlapping degree quite
likely will have a number of different possibilities to choose the
first edge to go towards the other nodes, and in this way it will have
a low value of $\lambda_i$. Conversely, a node with low degree will
more likely have a high value of $\lambda_i$, being its shortest paths
constrained to a limited selection of edges and layers from the first
step. Moreover, $\lambda_i$ appears to be slightly anti-correlated with
$P_i$, as confirmed by the values of Kendall's and Pearson's
correlation coefficients (see the caption of Fig.~\ref{fig:figure7}).
\begin{figure*}[!htbp]
\begin{center}
\includegraphics[width=6in]{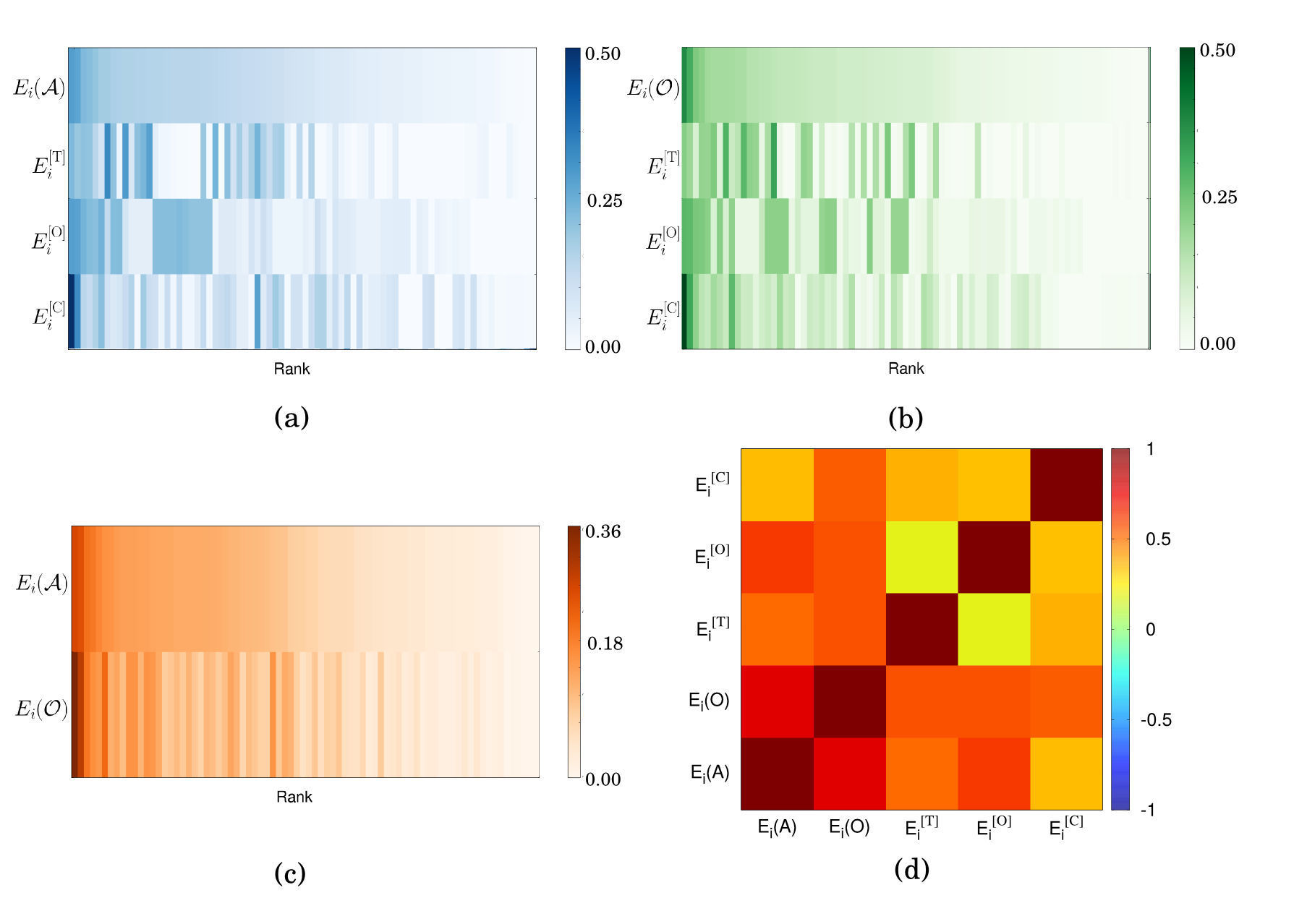}
\caption[]{(color online) (a) Eigenvector centrality of the aggregated
  topological network $E_i(\mathcal A)$, and of the trust $E_i^{[\rm
      T]}$, operational $E_i^{[\rm O]}$ and communication layer
  $E_i^{[\rm C]}$.  The nodes are ranked according to their value of
  $E_i(\mathcal A)$ on the aggregated topological network. (b) Similar
  to panel (a) but here nodes are ranked according to their
  eigenvector centrality computed on the aggregated overlapping
  network $E_i(\mathcal O)$. (c) Comparison of the rankings of
  eigenvector centrality computed on the aggregated topological
  network and on the aggregated overlapping network, respectively
  $E_i(\mathcal A)$ and $E_i(\mathcal O)$. (d) The heat map shows the
  non-parametric correlation between the rankings induces by the
  different centralities.}
\label{fig:figure8}
\end{center}
\end{figure*}

\section{\textbf{CENTRALITY}}
\label{section:centrality}

The concept of node centrality is an important and well-studied issue
in network theory. Various measures of centrality, such as the node
degree, the closeness and the betweenness, have been proposed and used
over the years to quantify the importance of a node in a single-layer
network~\cite{wasserman94}. The extension of these concepts to
multiplex networks is still an open research question. In
Section~\ref{section:formalism} we have proposed various ways to
extend the definition of node degree to the case of a multi-layer
system. Here, we will focus our attention on the eigenvector
centrality, which is a generalization of the concept of degree
centrality.  In a single-layer network the eigenvector centrality of a
node $i$ is defined as the $i$-th component of the eigenvector
associated to the leading eigenvalue of the adjacency matrix of the
network~\cite{bonacich}. For a multiplex network, we can calculate the
eigenvector centrality at each layer. If we denote as
$E_i^{[\alpha]}$, the eigenvector centrality of node $i$ at layer
$\alpha$, then the eigenvector centrality of node $i$ in the multiplex
network is a vector:
\begin{equation}
\bm{E}_i =\{E_i^{[1]}, \ldots, E_i^{[M]}\}.
\end{equation}
We can also compute the eigenvector centrality on the aggregated
topological and on the aggregated overlapping network. We indicate the
results respectively as $E_i(\mathcal A)$ and $E_i(\mathcal O)$. In
Fig.~\ref{fig:figure8}(a), \ref{fig:figure8}(b) and
\ref{fig:figure8}(c) we compare the eigenvector centrality computed on
each layer with that evaluated on the aggregated topological and
overlapping networks. We notice only very weak correlations between
the different centrality sequences.  Such results are very similar to
those obtained in Section~\ref{section:node} for the case of node
degree, as a consequence of the fact that, at order zero, the
eigenvector centrality reduces to the node degree. The Kendall
correlation coefficients obtained for pairs of centralities are
reported in Fig.~\ref{fig:figure8}(d) as a heat map.
For a large fraction of nodes, the rankings induced by the eigenvector
centrality at different layers differ significantly. A slightly higher
value of correlation is found between centrality at different layers
and the centrality of the aggregated network, while the maximum
correlation is observed between the values of eigenvector centrality
computed on the aggregated topological network and on the aggregated
overlapping network.

\begin{figure}
\begin{center}
\includegraphics[width=3.1in]{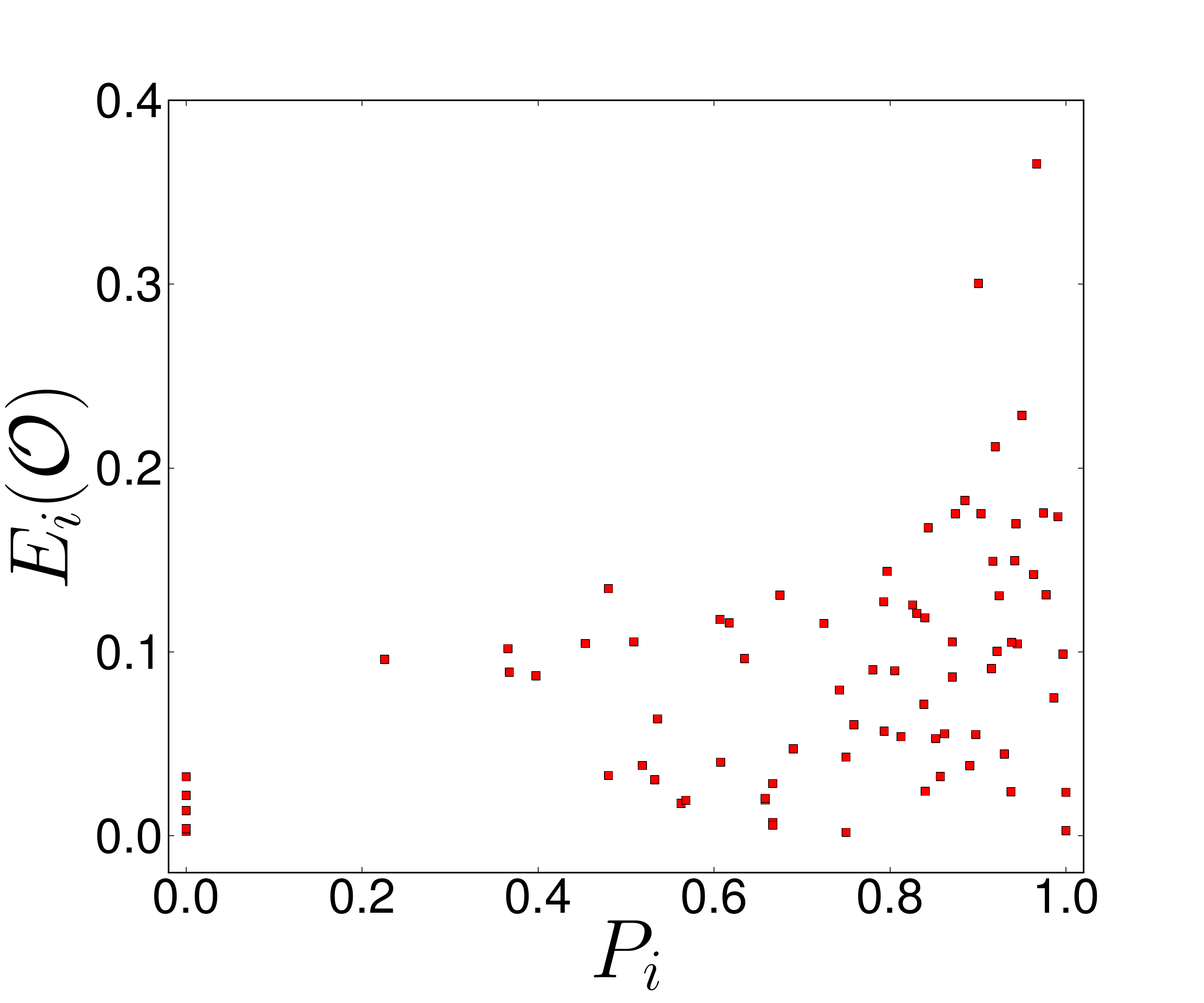}
\caption[]{Scatter-plot of the eigenvector centrality on the
  aggregated overlapping network $E_i(\mathcal O)$, and the
  participation coefficient $P_i$. Notice that there is indeed a
  slightly positive correlation between these metrics
  (${\tau(E_i(\mathcal O),P_i)=0.31}$, ${r(E_i(\mathcal
    O),P_i)=0.43}$).}
\label{fig:figure9}
\end{center}
\end{figure}

It is interesting to notice, as shown in Fig.~\ref{fig:figure9}, that
the centrality computed on the aggregated networks (e.g., on the
overlapping network) is not correlated with the multiplex
participation coefficient of the nodes. In fact, if we fix the value
of $E_i(\mathcal O)$, we observe a large heterogeneity in the values
of $P_i$, and vice-versa.

Until now we have computed and compared the eigenvector centralities
at each layer of the network. As already done for other metrics, we
will now propose a proper multiplex definition of the eigenvector
centrality which takes into account the presence of all layers at the
same time.  We follow a similar but relatively simpler approach than
the one recently proposed in~\cite{eigenvector}.  Given a two-layer
multiplex network (a duplex) and the corresponding adjacency matrices
$A^{[1]}$ and $A^{[2]}$, we can construct the following adjacency
matrix:
\begin{equation}
\mathcal M(b)=bA^{[1]}+(1-b)A^{[2]},
\label{eqcentrality1}
\end{equation} 
which is a convex combination of $A^{[1]}$ and $A^{[2]}$ where $b$ is
a parameter taking values in the interval $[0,1]$. We call such matrix
the multi-adjacency matrix. Notice that the parameter $b$ sets the
relative contribution of each layer to the multiplex structure. In
fact, if $b=0$ (respectively, $b=1$) the multi-adjacency matrix of the
duplex reduces to $A\lay{2}$ (respectively $A\lay{1}$). We can
consider $b=0.5$ as the benchmark case, where the two layers are given
the same weight. Notably, we have ${\mathcal M(b=0.5)= \mathcal O/2}$,
i.e. for $b=0.5$ the multi-adjacency matrix is proportional to the
aggregated overlapping network.

\begin{figure}
\begin{center}
\includegraphics[scale=0.3]{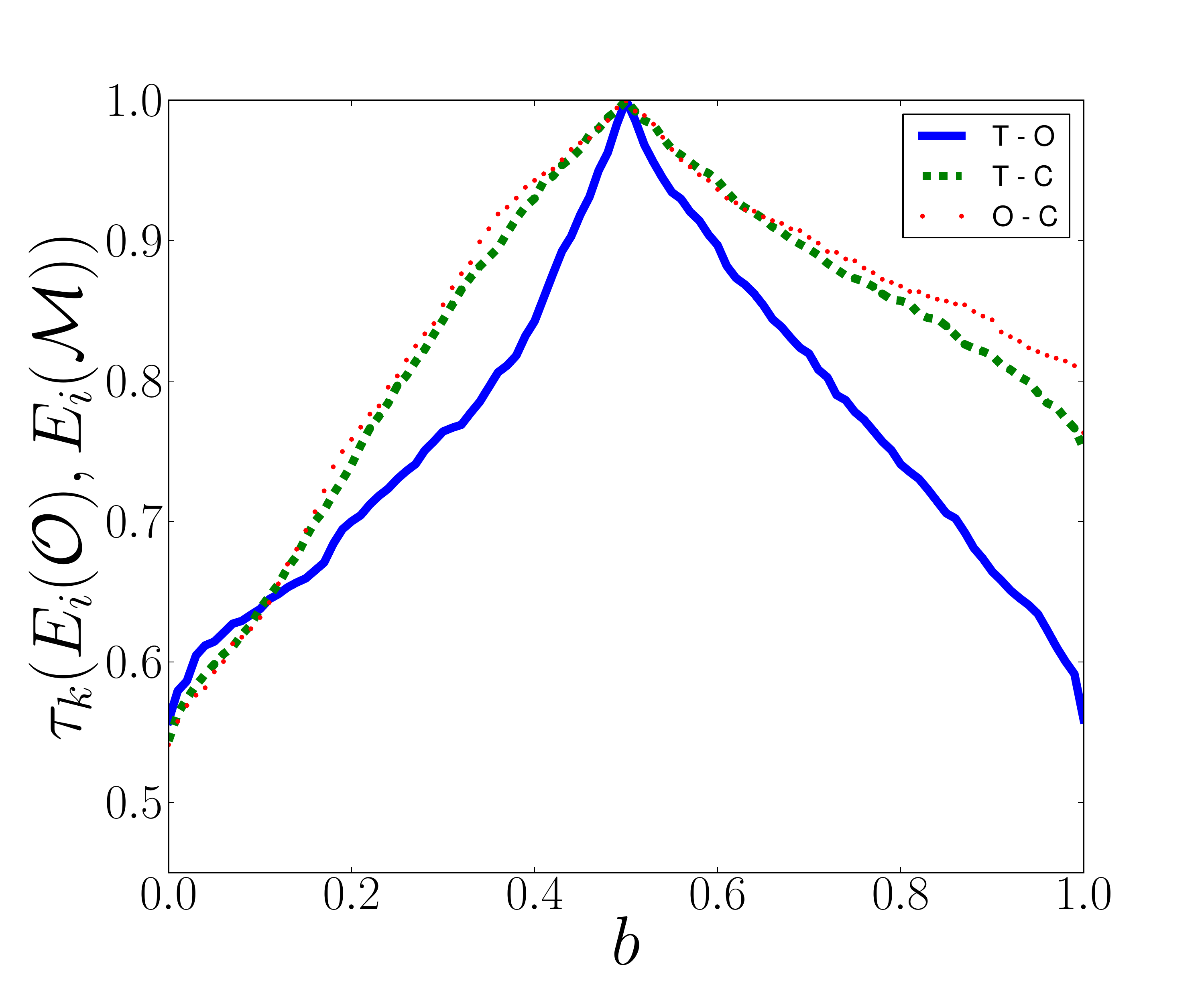}
\caption[]{(color online) For each couple of layers (duplex) of the multiplex network
  of Indonesian terrorists we plot the Kendall correlation coefficient
  $\tau_k$ between the eigenvector centrality of the benchmark case
  ($b=0.5$, i.e. equal weights on both layers) and the generic case of
  Eq.~\ref{eqcentrality1}. }
\label{centrality1}
\end{center}
\end{figure}

For each value of $b$, $\mathcal M$ is a square matrix with
non-negative entries. Thus, being satisfied all the hypotheses of the
Perron-Frobenius theorem, we can calculate the eigenvector centrality
of $\mathcal M$ as a function of $b$. In order to assess the role of
each layer in determining the multiplex centrality, we follow this
approach: we compute the eigenvector centrality of the benchmark case
$b=0.5$ (corresponding to matrix $\mathcal O$); we then compute the
eigenvector centrality of $\mathcal M$ for a generic value of
$\overline{b}$, and we evaluate the Kendall correlation coefficient
$\tau_k$ between the centrality ranking obtained for $b=\overline{b}$
and the benchmark case $b=0.5$.  Since the multiplex network of the
Indonesian terrorists has three layers, we can construct three
different duplex networks. The results are shown in
Fig.~\ref{centrality1}, where we plot the Kendall coefficient $\tau_k$
as a function of $b$.

As expected, the three duplex have a peak $\tau_k=1$ for $b=0.5$.  By
comparing the three curves we can deduce that T and O have a similar
role in determining the centrality of the multi-layer system, in both
cases stronger than layer C. In fact, the slopes of the curves, as
well as their symmetry/asymmetry, and the symmetry/asymmetry of the
extreme cases $b=0$ and $b=1$, tell us about the interplay between the
two layers in determining the centrality of the multi-layer system.
The curve corresponding to the duplex T-O is quite symmetrical,
indicating that the effect of T and O on the centrality is very
similar. Conversely, the curves corresponding to T-C and O-C are
asymmetrical. This means that both layers T and O dominate layer C in
determining the centrality of the nodes.  If we focus on the case
$b=0$, we obtain three similar values of $\tau_k$. Instead, the three
curves display different behavior in the range $0 \le b \le 0.5$. In
particular, the solid blue curve shows the steepest decrease from the
peak (this is also true for $b \ge 0.5$), indicating that layers T and
O are more different than layers T and C or layers O and C. For this
reason, a small perturbation of the coefficients of $\mathcal M$ from
the benchmark case affects the centrality of the multi-layer system
more for the duplex T-O than for the duplexes T-C and O-C. The largest
dissimilarity of the pair T-O is also confirmed by the smallest value
of $\tau_k$ found for the couple $E_i^{[\rm T]}$ and $E_i^{[\rm O]}$,
as shown in Fig.~\ref{fig:figure8}(d).
\begin{figure}[t]
\begin{center}
  \includegraphics[width=3in]{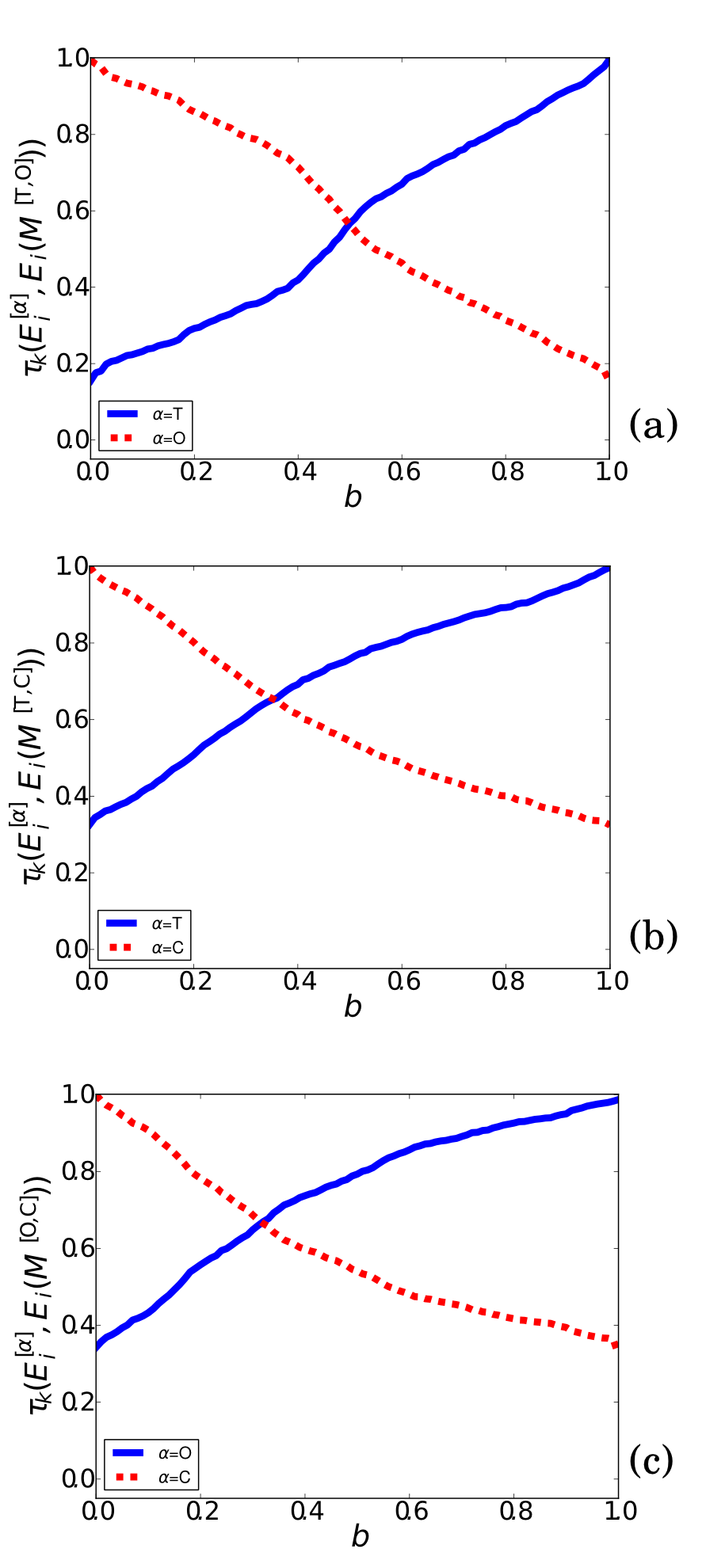}
  \caption[]{(color online) As subsets of the original overlapping network, we consider the three duplex $M^{[\rm T, \rm O]}=bA^{\rm T}+(1-b)A^{[\rm O]}$, $M^{[\rm T, \rm C]}=bA^{\rm T}+(1-b)A^{[\rm C]}$ and $M^{[\rm O, \rm C]}=bA^{\rm O}+(1-b)A^{[\rm C]}$. For each possible duplex, we report the Kendall coefficient $\tau_k$ between the centrality of each single
    layer and the corresponding $\mathcal M$ as a function of $b$.}
  \label{centrality2}
\end{center}
\end{figure}

A slightly different approach provides useful insights about the
distribution of centrality in the system under study. Given the three
duplex networks, for each one of them we can compute the Kendall
coefficient $\tau_k$ between the values of centrality obtained for
$\mathcal M$ and different values of $b$, and those obtained for each
single layer.  Results are shown in Fig.~\ref{centrality2}. We note
that the value of $\tau_k(E_i^{[\alpha]}, E_i(\mathcal M(b=0.5)))$ in
each panel of Fig.~\ref{centrality2} is equal, respectively, to the
value of $\tau_k(E_i(\mathcal O),E_i(\mathcal M(b=1)))$ for $\alpha=1$
and to $\tau_k(E_i(\mathcal O),E_i(\mathcal M(b=0)))$ for $\alpha=2$
on the corresponding curve in Fig.~\ref{centrality1}. In
Fig.~\ref{centrality2}(a) the two curves are quite symmetrical and
intersect around $b=0.5$, indicating that the contributions of layers
T and O to centrality is similar. Conversely, for both T-C and O-C
(respectively, Fig.~\ref{centrality2}(b) and
Fig.~\ref{centrality2}(c)) the two curves are asymmetrical and
intersect at $0.35 < b < 0.40$, indicating that both layer T and O
have stronger impact on centrality than layer C.  

These results indicate that multi-layer systems are characterised by
non-trivial organisation also with respect to centrality. We conclude
this Section by noticing that the definition of multiplex centrality
can be easily generalised the to a system of $M$ levels by
constructing the adjacency matrix:
\begin{equation}
\mathcal M = b_1A^{[1]} + b_2 A^{[2]} + ... + b_M A^{[M]}
\end{equation}
with the condition that $\sum_{i=1}^{M} b_i = 1$. Once again the
benchmark case obtained by fixing $b_1 = ... = b_{M} = \frac{1}{M}$
coincides with the aggregated overlapping network.

\section{\textbf{CONCLUSIONS}}
\label{section:conclusions}
The basic units of many real world systems are connected through a
large variety of different relations. One of the new challenges in
network theory is therefore to treat together ties of different kind
preserving existing differences. The multiplex metaphor, which allows
to distinguish the different kinds of relationships among a set of
nodes, constitutes a promising framework to study and model
multi-layer systems.  In this paper we proposed a comprehensive
formalism to deal with systems composed of several layers, both with
binary or weighted links. In particular, we provided a clear
distinction about the different levels of description of a multiplex
network: the aggregated topological, the overlapping and the weighted
overlapping network, which are simpler but less rich structures than
the vector of adjacency matrix $\bm{A}$.  We also proposed a number of
metrics to characterize multiplex systems with respect to node degree,
edge overlap, node participation to different layers, clustering
coefficient, reachability and eigenvector centrality. All these
measures were tested on the multiplex network of Indonesian
terrorists, a system with $78$ nodes and four layers.
Admittedly, the notation proposed in this work, based
  on the explicit vectorial representation of node and edge
  properties, is just one of the possible ways of dealing with
  multiplex networks, and indeed there have been other recent attempts
  to define a consistent framework for the analysis and
  characterization of multi-layer systems. In particular, the
  tensorial formalism proposed in Refs.~\cite{DeDomenico2013,arenas2}
  seems a promising approach, since it allows to express some
  multiplex metrics in a synthetic and compact way. However, we
  believe that the notation we proposed here, which makes explicit
  the role of single layers, is somehow more immediate
  to understand and easier to use for the study of real-world
  multiplex networks.  We really hope that the set of tools and
metrics presented in this paper will trigger further research on the
characterization of the structural properties of multi-layer complex
systems.

\begin{acknowledgements}
This work was supported by the Project LASAGNE, Contract No.318132 (STREP), funded by the European Commission. 
\end{acknowledgements}

\clearpage

%BIBLIOGRAFIA


\begin{thebibliography}{99}
\bibitem{note_undirected} Throughout the paper we make the assumption
  that the graph at each layer is simple, i.e. without self-loops.
\end{thebibliography}
\end {document}